\documentclass[a4paper,11pt]{article}

\usepackage[left=2.0cm,right=2.0cm,top=2.0cm,bottom=2.0cm]{geometry}

\usepackage{lineno}
\usepackage{abstract}
\usepackage{orcidlink}

\usepackage[utf8]{inputenc}
\pdfoutput=1
\usepackage{graphicx}
\usepackage{amsmath}
\usepackage{amsfonts}
\usepackage{amssymb}
\usepackage{xcolor,soul}
\usepackage{epstopdf}
\usepackage{xcolor}
\usepackage{csquotes}

\usepackage{float}
\usepackage{cite}

\usepackage{relsize}
\usepackage{graphics}
\usepackage{hyperref}
\usepackage{mathrsfs}
\usepackage{amssymb}
\usepackage{booktabs}
\usepackage[normalem]{ulem}
\usepackage{amsthm}
\usepackage{cancel}
\usepackage{fontawesome}   
\usepackage{tabularx,ragged2e}

\usepackage{tikz}
\usetikzlibrary{positioning,decorations.pathmorphing}

\hypersetup{
     colorlinks   = true,
     citecolor    = blue,
     linkcolor    = blue,
     urlcolor     = blue,
}
\usepackage{physics}

\usepackage{caption}
\usepackage{subcaption}

\title{Inflaton Accretion onto Primordial Black Holes During Reheating}

\author{Jitumani Kalita \thanks{k.jitumani@iitg.ac.in} \,\orcidlink{0009-0001-5681-6194} and Debaprasad Maity \thanks{debu@iitg.ac.in} \,\orcidlink{0000-0002-5458-7121} \\
Department of Physics, Indian Institute of Technology Guwahati,\\
Assam, India}

\date{}

\newcommand{\bea}{\begin{aligned}}
\newcommand{\eea}{\end{aligned}}

\def\beq{\begin{equation}}
\def\eeq{\end{equation}}

\def\beqa{\begin{eqnarray}}
\def\eeqa{\end{eqnarray}}

\def\be{\begin{equation}}
\def\ee{\end{equation}}

\def\bse{\begin{subequations}}
\def\ese{\end{subequations}}

\def\bea{\begin{eqnarray}}
\def\eea{\end{eqnarray}}

\newcommand{\mpl}{M_{p}}

\usepackage{pgfplots}
\pgfplotsset{compat=1.17}

\begin{document}

\maketitle



\begin{abstract}
    Primordial Black Holes (PBHs) forming prior to Big Bang Nucleosynthesis evolve during the reheating epoch, an environment dominated by an oscillating inflaton field decaying into a relativistic thermal bath. In this work, we track the complete lifecycle of PBHs within this coupled inflaton-radiation background. Utilizing $\alpha$-attractor E-models, we analytically anchor the reheating initial conditions directly to Cosmic Microwave Background observations. By matching exact scalar field solutions in a Schwarzschild spacetime to the cosmological far-zone, we derive the cycle-averaged mass accretion rate and couple it to the growing radiation bath. We find that this combined accretion induces a highly non-linear enhancement of the final PBH mass. Because the Hawking evaporation timescale scales cubically with mass, PBHs forming near their critical runaway limits experience a massive extension of their lifespans. Surviving deeper into the radiation-dominated era triggers a multi-order-of-magnitude amplification in their emitted Stochastic Gravitational Wave Background (SGWB).
\end{abstract}

\section{Introduction}
\label{sec:introduction}

Primordial Black Holes (PBHs) are unique cosmological relics that offer a profound window into the physics of the early Universe. First proposed in the 1970s \cite{Zeldovich:1967lct, Hawking:1971ei, Carr:1974}, PBHs are thought to form from the gravitational collapse of extreme primordial density fluctuations~\cite{Carr:1975qj, Chapline:1975ojl}. Unlike astrophysical black holes (BHs), which are born from stellar collapse late in cosmic history, PBHs could have formed essentially at any epoch prior to Big Bang Nucleosynthesis (BBN), spanning a vast spectrum of possible masses~\cite{Carr:1975qj, Keith:2020jww, Carr:2020gox}. Consequently, they remain one of the most compelling candidates for dark matter~\cite{Carr:2016drx, Carr:2020xqk, Green:2020jor, Belotsky:2014kca}, as well as potential seeds for supermassive BHs~\cite{Bean:2002kx, Carr:2018rid, Sasaki:2018dmp, Khlopov:2008qy} and sources of exotic gravitational wave events~\cite{Bird:2016dcv, Clesse:2016vqa}.

The environment of the early Universe dictates both the formation threshold and the subsequent evolution of a PBH. Standard treatments often assume that PBHs form and evolve in a universe dominated by a perfect radiation fluid~\cite{Carr:1975qj, Niemeyer:1997mt, Musco:2004ak, Harada:2013epa}. However, this assumption bypasses the crucial physics of the reheating epoch, the transitional phase bridging the end of cosmic inflation and the onset of the thermalized radiation-dominated era \cite{Kofman:1997yn, Bassett:2005xm, Allahverdi:2010xz, Allahverdi:2020bys}. During reheating, the Universe's energy budget is known to be dominated by the coherent oscillations of the inflaton scalar field about the minimum of its potential~\cite{Turner:1983he, Kofman:1997yn, Bassett:2005xm}. The macroscopic behavior of this oscillating field, including its effective equation of state, is strictly governed by the shape of the potential, commonly modeled as $V(\Phi) \propto \Phi^{2n}$~\cite{Turner:1983he, Johnson:2008se, Garcia:2020wiy, Ando:2017veq, Carr:2018nkm}.

A PBH born in this dynamic environment will not remain static. It will accrete the surrounding background energy density~\cite{Zeldovich:1967lct, Carr:1974, Harada:2004pf, Kallifatides:2026sik}. Although the accretion of standard cosmological fluids (like matter or radiation) has been widely studied~\cite{Bondi:1952ni, Michel1972ApSS, Custodio:2002gj, Lora-Clavijo:2013aya, Mack:2006gz, Ali-Haimoud:2016mbv}, the accretion of an oscillating scalar field in an expanding background presents a highly non-trivial boundary value problem~\cite{Harada:2004pf, Babichev:2004yx, Padilla:2021zgm}. Furthermore, because the inflaton field is continuously decaying to the standard model thermal bath, the PBH is subjected to a two-component accretion process: it accrets both the oscillating inflaton and the newly born relativistic radiation.

Because the ultimate fate of a PBH is governed by Hawking evaporation \cite{Hawking:1974rv, Hawking:1975vcx}, accurately determining its fully accreted mass is of paramount importance. The evaporation timescale scales with the BH mass as ($t_{\rm ev} \propto M^3$). Therefore, even a seemingly brief phase of rapid accretion during the reheating epoch can significantly enhance the PBH's lifespan, determining whether it evaporates completely during the cosmological evolution or survives long enough to leave direct observable consequences. For PBHs that evaporate completely before BBN, the most robust observable signature is the emission of a stochastic gravitational wave background (SGWB) \cite{Anantua:2008am, Dong:2015yjs, Romano:2016dpx, Caprini:2018mtu, Yuan:2021xdi, Gross:2024wkl, Hooper:2019gtx, Dolgov:2011cq, Masina:2020xhk, Jia:2026psi}. As the PBH shrinks, its Hawking temperature diverges, leading to an explosive emission of all fundamental particles, including gravitons. These high-frequency gravitational waves act as a form of dark radiation. The integrated energy density of those gravitons is strictly bounded by measurements of the effective number of relativistic degrees of freedom ($\Delta N_{\rm eff}$) from the Cosmic Microwave Background (CMB)~\cite{Hooper:2019gtx, Arbey:2021ysg, Das:2023oph, Planck:2018vyg}.

In this work, we provide a comprehensive analytical and numerical framework for tracking the complete life cycle of a PBH formed during the reheating epoch. We model the inflationary sector using the observationally favored $\alpha$-attractor E-models. By matching the microphysics of the potential directly to the latest CMB observables (the scalar spectral index $n_s$ and the tensor-to-scalar ratio $r$), we rigorously anchor the initial conditions of the reheating epoch. We then solve the Klein-Gordon equation in the Schwarzschild background living in an expanding background to derive the analytic accretion rate of the oscillating inflaton field, and seamlessly couple this to the accretion of the growing radiation bath.
Important to note that during reheating, the Inflaton field is usually described by a fluid with an effective equation of state. However, in the ultralight black hole background, it is the oscillating field description that captures the correct PBH accretion process as opposed to effective fluid description of the inlaton field. We show that indeed the PBH accretion is less efficient for inflaton as an oscillating field than as a fluid with an effective equation of state.  

Such distinct behavior will be shown to emerge for those PBHs that transit around the end of the reheating phase while accreting. 
Around this period the PBHs environment dynamically changes from inflaton field to radiation fluid. 
Therefore, if the PBH accretion process is ongoing during the end of reheating, it receives a sudden kick due to enhanced radiation production. Finally, we trace the non-linear Hawking evaporation phase and compute the resulting present-day SGWB spectrum, confronting our results with $\Delta N_{\rm eff}$ bounds from Planck~\cite{Planck:2018vyg} and future CMB-S4~\cite{Abazajian:2019eic} projections. Most important result of our present analysis is the dependence of the GW spectrum on the reheating temperature.

The paper is organized as follows. In Section \ref{sec:flrw_background}, we study the inflaton dynamics in an expanding FLRW background for various potential shapes and track the coupled inflaton--radiation system during reheating. We then relate the reheating duration and temperature to CMB observables. In Sections \ref{sec:bh_spacetime} and \ref{sec:matching_accretion}, we evaluate the scalar field in the BH spacetime and match the solutions to derive the mass accretion rate. Sections \ref{sec:hawking_evaporation} and \ref{sec:gravitational_waves} track the subsequent Hawking evaporation and the generation of the SGWB. Finally, we conclude in Section \ref{sec:conclusion}..

\section{Evolution of the Inflaton Field during Reheating}
\label{sec:flrw_background}


We begin by establishing the dynamics of the scalar field in the cosmological background which can be identified as a far-zone from the centre of a PBH, which is governed by the background Friedmann-Lema\^{i}tre-Robertson-Walker (FLRW) expansion. We model the post-inflationary epoch as being dominated by the coherent oscillations of the inflaton field about the minimum of its potential. This oscillating field behaves as a fluid with a specific equation of state, driving the expansion of the Universe.

We consider a general Lagrangian density for a minimally coupled scalar field $\Phi$ with a potential $V(\Phi)$ 
\begin{equation}
    \mathcal{L} = -\frac{1}{2}g^{\mu\nu}(\partial_{\mu}\Phi)(\partial_{\nu}\Phi) - V(\Phi).
\end{equation}
During the reheating phase, the inflaton oscillates around the minimum of its potential, which is typically well-approximated by a monomial form $V(\Phi) \propto \Phi^{2n}$. This form arises naturally from the expansion of more fundamental inflationary models, such as $\alpha$-attractor potentials, around their minima ($\Phi=0$). For instance, the E-model potential is given by~\cite{Kallosh:2013yoa, Kallosh:2013lkr, Kallosh:2013tua, Galante:2014ifa}
\begin{equation}\label{eq:Potential_E_model}
    V(\Phi) = \Lambda^4\left(1 - e^{-\sqrt{\frac{2}{3\alpha}}\frac{\Phi}{M_{p}}}\right)^{2n} ,
\end{equation}
where $M_{p} \equiv 1/\sqrt{8 \pi G} \simeq 2.435 \times 10^{18}\, \mathrm{GeV}$ is the reduced Planck mass~\cite{Planck:2018vyg}. The parameter $\Lambda$ sets the overall energy scale of inflation and determines the energy density stored in the inflaton field during the inflationary epoch. Its magnitude is constrained by CMB observations and is related to the scalar spectral index $n_s$, the amplitude of scalar perturbations $A_s = 2.1 \times 10^{-9}$, and the tensor-to-scalar ratio $r$~\cite{Planck:2018vyg, Planck:2018jri}. This class of models is well supported by recent combined observational analyses (Planck+ACT+DESI+BICEP/Keck), which yield 
$n_s = 0.9743 \pm 0.0034$ at $68\%$ C.L., along with the $95\%$ C.L. upper bound $r_{0.05} < 0.038$~\cite{AtacamaCosmologyTelescope:2025blo, AtacamaCosmologyTelescope:2025nti, DESI:2024mwx, BICEP:2021xfz}. Within this framework, the tensor-to-scalar ratio can be written analytically as~\cite{Kallosh:2013yoa, Kallosh:2013lkr}
\begin{equation}
r = \frac{192 \alpha n^2 (1 - n_s)^2}
{\left[ 4n + \sqrt{16n^2 + 24\alpha n (1 - n_s)(1 + n)} \right]^2}.
\end{equation}
Here, the parameter $\alpha$ controls the curvature of the potential and hence the inflationary dynamics. 
The inflationary energy scale, expressed through $\Lambda$, can likewise be written in terms of the CMB observables as~\cite{Drewes:2017fmn}
\begin{equation}
\Lambda = M_{p}
\left( \frac{3\pi^2 r A_s}{2} \right)^{\frac{1}{4}}
\left[
\frac{2n(1 + 2n) + \sqrt{4n^2 + 6\alpha (1 + n)(1 - n_s)}}
{4n(1 + n)}
\right]^{\frac{n}{2}}.
\end{equation}
In the $\Phi \rightarrow 0$ limit, the potential \eqref{eq:Potential_E_model} can be expanded as
\begin{equation}
V(\Phi\to 0) \simeq \Lambda^4\left(\frac{2}{3\alpha}\right)^n \left(\frac{\Phi}{M_{p}}\right)^{2n} .
\end{equation}

\subsection{The Oscillating Scalar Field as a Cosmological Fluid}
\label{subsec:scalar_fluid}

In a spatially flat FLRW background, $ds^2 = -dt^2 + a^2(t)d\mathbf{x}^2$, the equation of motion for a homogeneous scalar field ($\Phi(t)$) is the Klein-Gordon equation given by~\cite{Kolb:1990vq, LiddleLyth2000}
\begin{equation}
    \ddot{\Phi} + 3H\dot{\Phi} + V'(\Phi) = 0, \label{eq:Inflaton_EOM}
\end{equation}
where $H = \dot{a}/a$ is the Hubble parameter and a prime denotes a derivative with respect to $\Phi$. The energy density $\rho_\Phi$ and pressure $P_\Phi$ of the scalar field are given by the time-like components of its energy-momentum tensor, $T^{\mu\nu} = \partial^\mu\Phi\partial^\nu\Phi - g^{\mu\nu}\mathcal{L}$, as~\cite{Baumann:2009ds, Mukhanov:2005sc}
\begin{align}
    \rho_{\Phi} = \frac{1}{2}\dot{\Phi}^2 + V(\Phi), \quad
    P_{\Phi} = \frac{1}{2}\dot{\Phi}^2 - V(\Phi).
\end{align}
Since the field is oscillating rapidly on a timescale much shorter than the Hubble time ($T_\text{osc} \ll H^{-1}$), we can determine its effective equation of state by averaging over one oscillation cycle~\cite{Turner:1983he}. Using the Virial theorem for a potential $V(\Phi) \propto \Phi^{2n}$, we have the relation $\langle \frac{1}{2}\dot{\Phi}^2 \rangle = n \langle V(\Phi) \rangle$~\cite{Turner:1983he, Johnson:2008se}. The cycle-averaged energy density and pressure are then become
\begin{equation}
\begin{aligned}
    \langle\rho_{\Phi}\rangle &= \langle \tfrac{1}{2}\dot{\Phi}^2 \rangle + \langle V(\Phi) \rangle = (n+1)\langle V(\Phi) \rangle, \\
    \langle P_{\Phi}\rangle &= \langle \tfrac{1}{2}\dot{\Phi}^2 \rangle - \langle V(\Phi) \rangle = (n-1)\langle V(\Phi) \rangle.
\end{aligned}
\end{equation}
This yields a constant, effective equation of state parameter, $\omega_\Phi$ as
\begin{equation}
    \omega_{\Phi} = \frac{\langle P_{\Phi} \rangle}{\langle \rho_{\Phi} \rangle} = \frac{n-1}{n+1}. \label{eq:eos}
\end{equation}
This simple relation maps the shape of the potential to the macroscopic behavior of the Universe. For a quadratic potential ($n=1$), the field behaves as pressureless matter ($\omega_\Phi = 0$). For a quartic potential ($n=2$), it behaves as radiation ($\omega_\Phi = 1/3$)~\cite{Turner:1983he, Albrecht:1982mp}.

The reheating era commences immediately after the end of cosmic inflation. The initial conditions for this new epoch are therefore determined by the physics at the inflationary exit~\cite{Kofman:1997yn, Bassett:2005xm, Allahverdi:2010xz}. The dynamics of inflation are governed by the slow-roll parameters, $\epsilon_V$ and $\eta_V$, which are defined in terms of the potential $V(\Phi)$ as~\cite{liddle2000, Baumann:2009ds}
\begin{align}
    \epsilon_V \simeq \frac{\mpl^2}{2} \left( \frac{V^{\prime}}{V} \right)^2, \quad
    \eta_V \simeq \mpl^2 \frac{V^{\prime \prime}}{V}.
\end{align}
Inflation ends when the slow-roll approximation breaks down, which occurs when $\epsilon_V \approx 1$~\cite{LiddleLyth2000, Liddle:2003as}. We can use this condition to find the Hubble parameter at the end of inflation, $H_{\rm end}$. From the definition $\epsilon_V \simeq \frac{1}{2}\dot{\Phi}^2 / (\mpl^2 H^2)$, the condition $\epsilon_V=1$ implies $\frac{1}{2}\dot{\Phi}^2_{\rm end} = \mpl^2 H^2_{\rm end}$. Substituting this into the Friedmann equation gives the Hubble parameter at the end of inflation 
\begin{equation}
    H_{\rm end} = \sqrt{\frac{V_{\rm end}}{2\mpl^2}}. \label{eq:H_end}
\end{equation}
For the $\alpha$-attractor E-model introduced in Eq.~\eqref{eq:Potential_E_model}, we can solve $\epsilon_V(\Phi_{\rm end}) = 1$ to find the field value at the end of inflation~\cite{Cook:2015vqa, Shojaee:2020xyr, Garcia:2020eof, Haque:2025uri, Saha:2025yzf}
\begin{equation}
    \Phi_{\rm end} = \sqrt{\frac{3 \alpha}{2}} \mpl \ln \left( \frac{2n}{\sqrt{3\alpha}}+1 \right). \label{eq:phi_end}
\end{equation}
The potential at this field value, $V_{\rm end} = V(\Phi_{\rm end})$, is then
\begin{equation}
    V_{\rm end} = \Lambda^4 \left( \frac{2n}{2n+\sqrt{3\alpha}} \right)^{2n}. \label{eq:V_end}
\end{equation}
These values, $\Phi_{\rm end}$ and $H_{\rm end}$, set the initial conditions for the subsequent evolution of the oscillating scalar field, which we explore in the following section.

\subsection{Amplitude Damping and Cosmological Evolution}
\label{subsec:amplitude_damping}

The interaction with the expanding background causes the amplitude of the inflaton field oscillations, which we denote $\Phi_0(t)$, to decrease over time. Therefore, one can parameterize the general solution of the inflaton as a product of the decaying part $\Phi_0(t)$, and an oscillating part $\mathcal{P}(t)$ as $\Phi(t)=\Phi_0(t) \mathcal{P}(t)= \Phi_0(t) \sum_{j} \mathcal{P}^{(n)}_{j} \cos(j \nu_n t) $~\cite{Turner:1983he, Kofman:1997yn, Shtanov:1994ce}, where $\nu_n$ is the fundamental frequency of oscillation for the periodic inflaton potential Eq.~\eqref{eq:Potential_E_model}~\cite{Garcia:2020wiy} for the inflation potential of form $\Phi^{2n}$. The above fundamental frequency turns out to be $\nu_n$, and is expressed as
\begin{equation}\label{eq:omega_n}
\nu_n (t) = \left.\sqrt{\frac{\partial^2 V}{\partial\Phi^2}}\right|_{\Phi_{0}} \frac{\sqrt{n \pi }}{\sqrt{2n-1}}  \frac{\Gamma\left(\frac{1}{2n}+\frac{1}{2}\right)}{\Gamma\left(\frac{1}{2n}\right)}.
\end{equation}
The evolution of the cycle-averaged energy density is governed by the continuity equation $\langle\dot{\rho}_\Phi \rangle + 3H(1+\omega_\Phi)\langle\rho_\Phi \rangle  = 0$. Noting that $\langle \rho_\Phi \rangle = V(\Phi_0) \propto \Phi_0^{2n}$, we can solve for the evolution of the amplitude as~\cite{Turner:1983he}
\begin{equation}
    \dot{\Phi}_0 = -\frac{3H}{n+1}\Phi_0.
\end{equation}
Integrating this equation yields the dependence of the amplitude on the scale factor $a(t)$ as~\cite{Turner:1983he, Shtanov:1994ce}
\begin{equation}
    \Phi_0(t) = \Phi_{\text{end}}\left(\frac{a(t)}{a_{\text{end}}}\right)^{-\frac{3}{n+1}}, \label{eq:phi0_evolution}
\end{equation}
where $\Phi_{\text{end}}$ and $a_{\text{end}}$ are the field amplitude and scale factor at the end of inflation, respectively. The background evolution is determined by the Friedmann equation, $H^2 = \langle \rho_\Phi \rangle / (3\mpl^2)$. Using $\langle \rho_\Phi \rangle \propto \Phi_0^{2n}$, we find the evolution of the Hubble parameter as
\begin{equation}
    H(t) = H_{\text{end}}\left(\frac{a(t)}{a_{\text{end}}}\right)^{-\frac{3n}{n+1}}. \label{eq:H_evolution_avg}
\end{equation}
This corresponds to a scale factor evolution of $a(t) \propto t^{\frac{n+1}{3n}}$. While Eq.~\eqref{eq:H_evolution_avg} describes the averaged cosmological evolution, the rapid oscillations of the field introduce periodic perturbations to the Hubble parameter. A more detailed calculation, which accounts for these oscillating dynamics, yields a corrected Hubble parameter. The derivation is lengthy and we present the key steps in Appendix~\ref{app:hubble_derivation} and the result is~\cite{Chakraborty:2023lpr}
\begin{equation}
H(t) \approx H_{\text{end}}\left(\frac{a(t)}{a_{\text{end}}}\right)^{-\frac{3n}{n+1}} \left( 1 + \frac{\mathcal{P}\sqrt{6(1-\mathcal{P}^{2n})}}{2(n+1)} \frac{\Phi_0(t)}{\mpl} \right),
\label{eq:H_corrected}
\end{equation}
where $H_{\text{avg}}(t)$ is the average Hubble rate from Eq.~\eqref{eq:H_evolution_avg}, and $\mathcal{P}(t) = \Phi(t)/\Phi_0(t)$ is the rapidly oscillating part of the solution.

\subsection{Solutions for Specific Potential Shapes}
\label{subsec:specific_cases}

We now apply this general formalism to three benchmark cases corresponding to different potential shapes $V \propto \Phi^{2n}$.

\paragraph{Quadratic Potential ($n=1$, $\omega_\Phi = 0$):} For a quadratic potential, the oscillating field behaves like matter. The field amplitude and the scale factor evolve as
\begin{align}
    \Phi_0(t) = \Phi_{\text{end}}\left(\frac{a}{a_{\text{end}}}\right)^{-3/2}, \quad
    a(t) \propto t^{2/3}.
\end{align}
The full solution for the scalar field, including its oscillation, can be approximated by its dominant mode
\begin{equation}
    \Phi(t) \approx \Phi_{\text{end}}\left(\frac{t}{t_{\text{end}}}\right)^{-1} \cos(\nu_1 t),
\end{equation}
where for $n=1$, the frequency of oscillation is given by
\begin{equation}\label{omega_n_1}
\nu_1 = \frac{2}{\sqrt{3\alpha}}\left(\frac{\Lambda}{\mpl}\right)^2 \mpl .
\end{equation}
Where ${\cal P}^{(1)}_1 = 1$, and all the higher harmonics are zero. The Hubble parameter's evolution, including the leading oscillatory correction from Eq. \eqref{eq:H_corrected}, becomes
\begin{equation}
    H(t) \approx H_{\text{end}}\left(\frac{t}{t_{\text{end}}}\right)^{-1}\left(1+\frac{\sin(2\nu_1 t)}{2\nu_1 t}\right).
\end{equation}

\paragraph{Quartic Potential ($n=2$, $\omega_\Phi = 1/3$):} For a quartic potential, the field behaves as radiation; this is a particularly well-motivated scenario for the reheating era. The amplitude and scale factor evolve as
\begin{equation}
    \Phi_0(t) = \Phi_{\text{end}}\left(\frac{a}{a_{\text{end}}}\right)^{-1}, \qquad
    a(t) \propto t^{1/2}.
\end{equation}
The field solution is a superposition of cosine modes, dominated by the first two non-zero terms 
\begin{equation}
    \Phi(t) \approx \Phi_{\text{end}}\left(\frac{t}{t_{\text{end}}}\right)^{-1/2} \left[ \mathcal{P}^{(2)}_1 \cos(\nu_2 t) + \mathcal{P}^{(2)}_3 \cos(3\nu_2 t) \right],
\end{equation}
where numerical calculations show $\mathcal{P}^{(2)}_1 \approx 0.47$ and $\mathcal{P}^{(2)}_3 \approx 0.02$~\cite{Chakraborty:2023lpr}. For the purpose of analytically matching assoicated the solution in the PBH background, we neglect the second harmonic correction since $\mathcal{P}^{(2)}_3 \ll \mathcal{P}^{(2)}_1$. Therefore, we can approximate the solution as,
\begin{equation}
    \Phi(t) \approx \mathcal{P}^{(2)}_1 \Phi_{\text{end}}\left(\frac{t}{t_{\text{end}}}\right)^{-1/2}  \cos(\nu_2 t).
\end{equation}
The Hubble parameter evolves as
\begin{equation}
    H(t) \approx H_{\text{end}}\left(\frac{t}{t_{\text{end}}}\right)^{-1}\left[ 1+\frac{\mathcal{P}\sqrt{6(1-\mathcal{P}^4)}}{6} \frac{\Phi_0(t)}{M_p}  \right].
\end{equation}

\begin{figure}[t]
\center
\includegraphics[width=\linewidth]{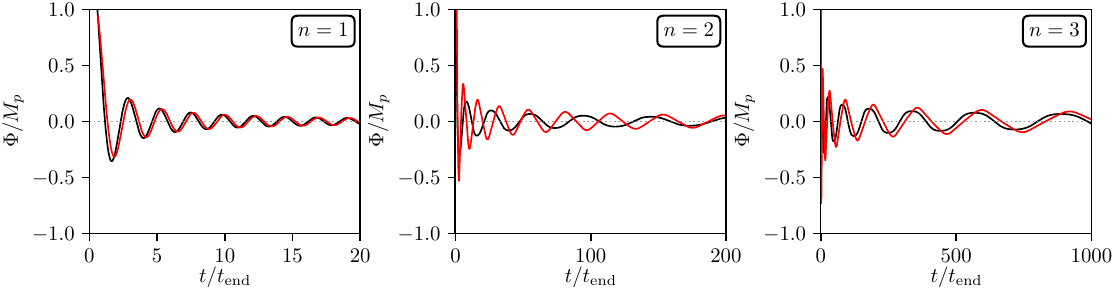} 
\caption{Oscillation of the scalar field $\Phi$ as a function of $t/t_{\rm end}$ for $n=1,2,3$, comparing the numerical results (red) with the analytical solutions (black).}
\label{fig:phi_evol}
\end{figure}

\paragraph{Sextic Potential ($n=3$, $\omega_\Phi = 1/2$):} For a sextic potential, the Universe is dominated by a fluid stiffer than radiation. The amplitude and scale factor evolve as
\begin{equation}
    \Phi_0(t) = \Phi_{\text{end}}\left(\frac{a}{a_{\text{end}}}\right)^{-3/4}, \qquad
    a(t) \propto t^{4/9}.
\end{equation}
Similar to the quartic case, the solution is a Fourier series with the dominant terms 
\begin{equation}
    \Phi(t) \approx \Phi_{\text{end}}\left(\frac{t}{t_{\text{end}}}\right)^{-1/3} \left[ \mathcal{P}^{(3)}_1 \cos(\nu_3 t) + \mathcal{P}^{(3)}_3 \cos(3\nu_3 t) \right],
\end{equation}
where $\mathcal{P}^{(3)}_1 \approx 0.46$ and $\mathcal{P}^{(3)}_3 \approx 0.03$~\cite{Chakraborty:2023lpr}. Because $\mathcal{P}^{(3)}_3 \ll \mathcal{P}^{(3)}_1$, the dominant contribution comes from the first harmonic, giving the approximate solution 
\begin{equation}
    \Phi(t) \approx \mathcal{P}^{(3)}_1 \Phi_{\text{end}}\left(\frac{t}{t_{\text{end}}}\right)^{-1/3} \cos(\nu_3 t).
\end{equation}
The Hubble parameter evolution is given by
\begin{equation}
    H(t) \approx H_{\text{end}}\left(\frac{t}{t_{\text{end}}}\right)^{-1}\left[ 1+\frac{\mathcal{P}\sqrt{6(1-\mathcal{P}^6)}}{8} \frac{\Phi_0(t)}{M_p} \right].
\end{equation}
We have plotted the analytical approximations along with the full numerical calculations of the time evolution of the scalar field $\Phi(t)$ as a function of $t/t_{\rm end}$ in Fig.~\ref{fig:phi_evol} for $n=1,2,3$. As seen in the left panel ($n=1$), the analytical solution governed by a simple harmonic oscillation is in excellent agreement with the numerical integration. For the higher-order potentials ($n=2$ and $n=3$, middle and right panels), the numerical results (red curves) exhibit additional high-frequency modulations. These modulations are a direct consequence of the anharmonic nature of the potential, which are absent in our analytical approximations (black curves) because we truncated the Fourier series to retain only the dominant harmonic ($\mathcal{P}^{(n)}_1$, $\mathcal{P}^{(n)}_3$), neglecting higher-order corrections. Nevertheless, the analytical expressions successfully track the macroscopic damping of the oscillation envelope across all three regimes, validating the derivation of the scale factor evolution and the background fluid approximation.


\subsection{Reheating Dynamics: Inflaton Decay and Radiation Production}
\label{sec:reheating_dynamics}

Throughout the post-inflationary epoch, the oscillating inflaton field does not merely dilute due to cosmic expansion; it continuously decays to populate the thermal bath of the standard model, ultimately transitioning the universe into the radiation-dominated era~\cite{Kofman:1997yn, Bassett:2005xm, Allahverdi:2010xz}. To accurately model the PBH accretion history during this period, we must track the energy density of both the inflaton field ($\rho_\Phi$) and the newly produced radiation ($\rho_r$). The evolution of these interacting fluids is governed by the coupled Boltzmann equations~\cite{Turner:1983he, Shtanov:1994ce, Kofman:1994rk, Giudice:2000ex, Haque:2023yra, Moss:2008lkw}
\begin{align}
\dot{\rho}_\Phi + 3H(1+\omega_\Phi)\rho_\Phi &= -\Gamma_{\Phi} \rho_\Phi (1+\omega_\Phi), \label{eq:boltz_phi} \\
\dot{\rho}_r + 4H\rho_r &= \Gamma_{\Phi} \rho_\Phi (1+\omega_\Phi), \label{eq:boltz_r}
\end{align}
where $\Gamma_\Phi$ is the perturbative decay rate of the inflaton field, which we treat as a free parameter associated with the underlying microphysics of the decay process~\cite{Bassett:2005xm, Garcia:2020mwi, Podolsky:2005bw}. The Hubble parameter evolves via the Friedmann equation, driven by the total energy density
\begin{equation}
H = \sqrt{\frac{\rho_\Phi + \rho_r}{3 M_p^2}}.
\end{equation}

\noindent
\textbf{Analytical Solutions during the Reheating Epoch:} During the early and intermediate stages of reheating, the decay rate is much smaller than the expansion rate ($\Gamma_\Phi \ll H$). The inflaton strictly dominates the energy budget ($\rho_\Phi \gg \rho_r$), allowing us to approximate $H^2 \simeq \rho_\Phi / 3M_p^2$. Under this approximation, the decay term in Eq.~\eqref{eq:boltz_phi} acts as a negligible perturbation to the inflaton's macroscopic evolution, yielding the standard dilution scaling~\cite{Kolb:1990vq, Turner:1983he}
\begin{equation}
\frac{\rho_\Phi}{\rho_\Phi^{\rm end}} \simeq \left( \frac{a}{a_{\rm end}} \right)^{-3(1+\omega_\Phi)}.
\end{equation}

Substituting this background evolution into the radiation equation \eqref{eq:boltz_r}, we can solve for the production of the radiation bath. Rewriting the left-hand side as $a^{-4} \frac{d}{dt}(\rho_r a^4)$ and converting the time derivative to a derivative with respect to the scale factor $a$ using $dt = da / (aH)$, we integrate from the end of inflation ($a_{\rm end}$) to an arbitrary scale factor $a$. Assuming the initial radiation density is negligible ($\rho_r^{\rm end} \simeq 0$), the integration yields~\cite{Cook:2015vqa, Dai:2014jja}
\begin{equation}
\frac{\rho_r}{\rho_\Phi^{\rm end}} \simeq \frac{2 (1+\omega_\Phi) }{(5-3\omega_\Phi)} \frac{\Gamma_\Phi}{ H_{\rm end}} \left( \frac{a}{a_{\rm end}} \right)^{-\frac{3}{2}(1+\omega_\Phi)}.
\label{eq:rho_r_evol}
\end{equation}
This indicates that the radiation energy density dilutes at a slower rate than the standard $a^{-4}$ scaling due to the continuous source term from the decaying inflaton.

\noindent
\textbf{Reheating Temperature and Scale Factor Evolution:} The reheating phase formally concludes when the radiation energy density overtakes the inflaton energy density, defining the scale factor at reheating, $a_{\rm re}$, where $\rho_r(a_{\rm re}) \simeq \rho_\Phi(a_{\rm re})$. Equating our analytical solutions, we find the expansion required to complete reheating
\begin{equation}
\frac{a_{\rm re}}{a_{\rm end}} \simeq \left[ \frac{(5-3\omega_\Phi)}{2(1+\omega_\Phi)} \frac{H_{\rm end}}{\Gamma_\Phi} \right]^{\frac{2}{3(1+\omega_\Phi)}}.
\end{equation}
At this juncture, the universe thermalizes at the reheating temperature, $T_{\rm re}$. The energy density of the radiation bath is given by $\rho_{\rm re} = (\pi^2 / 30) g_* T_{\rm re}^4$, where $g_*$ is the effective number of relativistic degrees of freedom~\cite{Kolb:1990vq, Baumann:2009ds}. Substituting the scaling relations at $a = a_{\rm re}$, the decay width $\Gamma_\Phi$ can be directly parameterized in terms of the reheating temperature~\cite{Cook:2015vqa, Dai:2014jja, Munoz:2014eqa}
\begin{equation}
\Gamma_\Phi \simeq \frac{(5-3\omega_\Phi)}{2(1+\omega_\Phi)} \left( \frac{\pi^2 g_*}{30} \right)^{\frac{1}{2}} \frac{T_{\rm re}^2}{\sqrt{3} M_p}.
\end{equation}
Finally, to map the PBH accretion history onto the reheating timeline, we must determine the total expansion from the moment of PBH formation ($a_{\rm in}$) to the end of reheating ($a_{\rm re}$). Using the relation $a \propto t^{\frac{2}{3(1+\omega_\Phi)}}$ and the initial formation time $t_{\rm in} = M_{\rm in} / (4 \pi \gamma M_p^2)$~\cite{Carr:1974, Carr:2016drx, Carr:2020xqk}, the ratio of the scale factors evaluates to
\begin{equation}
\frac{a_{\rm re}}{a_{\rm in}} = \left( \frac{a_{\rm re}}{a_{\rm end}} \right) \left( \frac{a_{\rm end}}{a_{\rm in}} \right) = \left\{ \frac{(5-3\omega_\Phi)}{2(1+\omega_\Phi)} \frac{4 \pi \gamma M_p^2}{\Gamma_\Phi M_{\rm in}} \right\}^{\frac{2}{3(1+\omega_\Phi)}}.
\label{eq:a_re_a_in}
\end{equation}
This geometric ratio fundamentally determines the duration over which the PBH is exposed to the inflaton and radiation environments, setting the stage for the combined accretion dynamics.


\subsection{Connecting the Reheating Epoch to CMB Observables}
\label{sec:cmb_matching}

To fully constrain the accretion history of PBHs formed during reheating, we must link the duration of the reheating phase and the thermalization temperature to observable cosmological parameters. Specifically, we can relate the reheating e-folds $N_{\rm re}$ and the reheating temperature $T_{\rm re}$ to the scalar spectral index $n_s$ and the tensor-to-scalar ratio $r$ measured at the cosmic microwave background (CMB) pivot scale $k$.

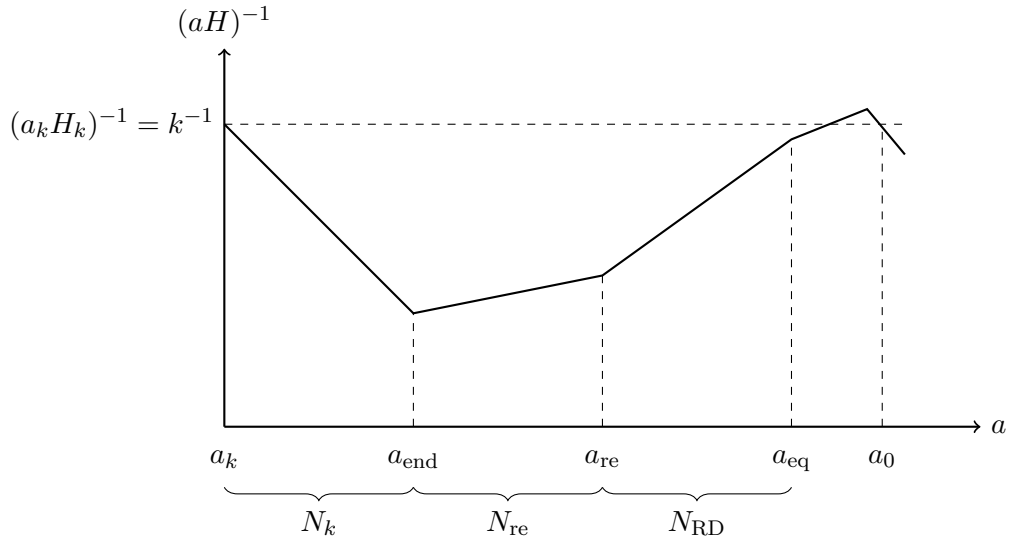
\begin{figure}[t]
\centering
\begin{tikzpicture}
    \draw[->, thick] (0,0) -- (10,0) node[right] {$a$};
    \draw[->, thick] (0,0) -- (0,5) node[above] {$(aH)^{-1}$};

    \draw[dashed] (0,4) node[left] {$(a_k H_k)^{-1} = k^{-1}$} -- (9,4);

    \draw[thick] (0,4) -- (2.5,1.5) -- (5,2) -- (7.5,3.8) -- (8.5,4.2) -- (9,3.6);

    \draw[dashed] (2.5,0) -- (2.5,1.5);
    \draw[dashed] (5,0) -- (5,2.0);
    \draw[dashed] (7.5,0) -- (7.5,3.8);
    \draw[dashed] (8.7,0) -- (8.7,3.9);
    
    \node[below=5pt] at (0,0) {$a_k$};
    \node[below=5pt] at (2.5,0) {$a_{\rm end}$};
    \node[below=5pt] at (5,0) {$a_{\rm re}$};
    \node[below=5pt] at (7.5,0) {$a_{\rm eq}$};
    \node[below=5pt] at (8.7,0) {$a_0$};

    \draw[decorate,decoration={brace,amplitude=5pt,mirror}] (0,-0.8) -- (2.5,-0.8) node[midway,below=5pt] {$N_k$};
    \draw[decorate,decoration={brace,amplitude=5pt,mirror}] (2.5,-0.8) -- (5,-0.8) node[midway,below=5pt] {$N_{\rm re}$};
    \draw[decorate,decoration={brace,amplitude=5pt,mirror}] (5,-0.8) -- (7.5,-0.8) node[midway,below=5pt] {$N_{\rm RD}$};
\end{tikzpicture}
\caption{Evolution of the comoving Hubble horizon $(aH)^{-1}$ with respect to the scale factor $a$, illustrating the horizon exit of the pivot scale $k$ and the subsequent cosmic epochs.}
\label{fig:comoving_horizon}
\end{figure}

As illustrated in Fig.~\ref{fig:comoving_horizon}, the physical mode of the CMB pivot scale $k$ crossed the comoving Hubble horizon during inflation ($k = a_k H_k$). We can relate this scale to the present-day horizon by tracking the expansion history across the subsequent cosmic epochs
\begin{equation}
\frac{k}{a_0 H_0} = \frac{a_k}{a_{\rm end}} \frac{a_{\rm end}}{a_{\rm re}} \frac{a_{\rm re}}{a_{\rm eq}} \frac{a_{\rm eq} H_{\rm eq}}{a_0 H_0} \frac{H_k}{H_{\rm eq}}.
\end{equation}
Taking the natural logarithm, this matching equation parameterizes the total expansion into the number of e-folds during inflation ($N_k$), reheating ($N_{\rm re}$), and radiation domination ($N_{\rm RD}$)
\begin{equation}
\ln\left( \frac{k}{a_0 H_0} \right) = - N_k - N_{\rm re} - N_{\rm RD} + \ln\left( \frac{a_{\rm eq} H_{\rm eq}}{a_0 H_0} \right) + \ln\left( \frac{H_k}{H_{\rm eq}} \right)
\label{eq:pivot_scale}
\end{equation}
where the Hubble parameter at the pivot scale is given in terms of the power spectrum amplitude $A_s$ and the tensor-to-scalar ratio $r$ as $H_k = \pi M_p \sqrt{r A_s/ 2}$.

To accurately compute the initial conditions for reheating, we must evaluate the slow-roll parameters directly from the specific shape of the $\alpha$-attractor E-model potential \eqref{eq:Potential_E_model}. During the slow-roll regime, the first and second Hubble flow parameters are given by $\epsilon_V \simeq \frac{M_p^2}{2} (V'/V)^2$ and $\eta_V \simeq M_p^2 (V''/V)$. Evaluating the derivatives of the E-model potential yields
\begin{equation}
\epsilon_V \simeq \frac{4n^2}{3\alpha \left(e^{\sqrt{\frac{2}{3\alpha}} \frac{\Phi}{M_p}} - 1 \right)^2}, \qquad \eta_V \simeq \frac{4n}{3\alpha} \frac{1}{\left(  e^{ \sqrt{\frac{2}{3\alpha}} \frac{\Phi}{M_p}} -1 \right)} \left[ \frac{2n-1}{ e^{\sqrt{\frac{2}{3\alpha}} \frac{\Phi}{M_p}} -1 } - 1 \right].
\end{equation}
The number of e-folds $N_k$ from the horizon exit of the pivot scale $\Phi_k$ to the end of inflation $\phi_{\rm end}$ is computed by integrating the slow-roll equation of motion, $N_k \simeq \int_{\Phi_k}^{\Phi_{\rm end}} [M_p \sqrt{2\epsilon_V^k}]^{-1} d\Phi$. Evaluating this integral analytically gives
\begin{equation}
N_k = \frac{\sqrt{6\alpha}}{4n M_p} (\Phi_{\rm end} - \Phi_k) - \frac{3\alpha}{4n} \left( e^{\sqrt{\frac{2}{3\alpha}} \frac{\Phi_{\rm end}}{M_p}} - e^{\sqrt{\frac{2}{3\alpha}} \frac{\Phi_k}{M_p}} \right).
\label{eq:Nk_exact}
\end{equation}
The primary CMB observables, the scalar spectral index ($n_s \simeq 1 - 6\epsilon_V^k + 2\eta_V^k$) and the tensor-to-scalar ratio ($r \simeq 16\epsilon_V^k$), are evaluated at $\Phi = \Phi_k$. By substituting our exact expressions for the slow-roll parameters into the spectral index equation, we can uniquely solve for the field value at horizon exit as a function of $n_s$
\begin{equation}
\Phi_k = \sqrt{\frac{3\alpha}{2}}\, M_p \ln \left(1 + \frac{4n + \sqrt{16n^2 + 24\alpha n (1 - n_s)(1 + n)}}{3\alpha (1 - n_s)}\right).
\label{eq:phi_k}
\end{equation}
Inserting this exact relation back into the expression for $\epsilon_V^k$, the tensor-to-scalar ratio is strictly determined by $n_s$, $\alpha$, and $n$
\begin{equation}
r = \frac{192 \alpha n^2 (1 - n_s)^2}{\left[ 4n + \sqrt{16n^2 + 24\alpha n (1 - n_s)(1 + n)} \right]^2}.
\end{equation}
With $\Phi_k$ fixed by the observable $n_s$, the inflationary e-folds $N_k$ are subsequently locked via Eq.~\eqref{eq:Nk_exact}, mapping the microphysics of the potential directly to the macroscopic expansion history. The field value at the end of inflation, $\Phi_{\rm end}$, is naturally defined where the slow-roll approximation breaks down ($\epsilon_V \simeq 1$).

The duration of reheating, $N_{\rm re} = \ln(a_{\rm re}/a_{\rm end})$, is intrinsically tied to the dilution of the inflaton energy density, $\rho_\Phi \propto a^{-3(1+\omega_\Phi)}$. By enforcing entropy conservation from the end of reheating to the present day ($g_{*s}(T_{\rm re}) T_{\rm re}^3 a_{\rm re}^3 = g_{*s}(T_0) T_0^3 a_0^3$), and utilizing the pivot scale matching established in Eq.~\eqref{eq:pivot_scale}, we can algebraically isolate the duration of reheating. After replacing $N_{\rm RD}$ via the horizon crossing conditions and the present-day photon temperature $T_0$, the exact analytical expression for the reheating e-folds becomes 
\begin{equation}
\begin{split}
N_{\rm re} &= \frac{4}{1-3\omega_\Phi} \left[ \frac{1}{4} \ln\left(\frac{\pi^2}{30} g_*\right) + \frac{1}{3} \ln\left(\frac{3.94}{g_{*s}(T_{\rm re})}\right) - \frac{1}{4} \ln\left(\frac{\rho_\Phi^{\rm end}}{T_0^4}\right) - N_k - \ln\left(\frac{k}{a_0 H_k}\right) \right],
\end{split}
\label{eq:N_re_analytic}
\end{equation}
where the energy density at the end of inflation is $\rho_\Phi^{\rm end} = 3M_p^2 H_{\rm end}^2$. With $N_{\rm re}$ determined, the reheating temperature is uniquely fixed by the corresponding thermalization relation 
\begin{equation}
T_{\rm re} = \left(\frac{\pi^2}{30} g_*\right)^{-\frac{1}{4}} \left(\rho_\Phi^{\rm end}\right)^{\frac{1}{4}} e^{-\frac{3}{4}(1+\omega_\Phi) N_{\rm re}}.
\label{eq:T_re_analytic}
\end{equation}
Equation \eqref{eq:N_re_analytic} reveals a critical physical singularity when the effective equation of state is $\omega_\Phi = 1/3$, which perfectly corresponds to an inflaton oscillating in a quartic potential ($V \propto \Phi^4$, $n=2$). In this specific scenario, the energy density of the coherent inflaton oscillations dilutes as $a^{-4}$ during reheating. This macroscopic behavior is completely indistinguishable from the dilution of a standard radiation bath. From the perspective of the CMB photons, the expansion history of the universe is identical whether the universe is filled with oscillating $\Phi^4$ inflaton particles or actual thermalized radiation. Because the expansion rate does not change when the inflaton finally decays, the CMB observables ($n_s$ and $r$) are completely ``blind'' to the epoch when reheating formally ended. 

Consequently, for the $n=2$ case, $T_{\rm re}$ is a free parameter. It is strictly unconstrained by $n_s$ and $r$, and can theoretically take any value between the Big Bang Nucleosynthesis bound ($T_{\rm BBN} \sim 4 \, \rm MeV$) and the maximum energy scale at the end of inflation ($\sim 10^{15}\, \rm GeV$). For models with $n \neq 2$, $T_{\rm re}$ remains definitively constrained by the CMB pivot scale.


\section{Scalar Field Dynamics in a Black Hole Spacetime}
\label{sec:bh_spacetime}

With the cosmological background established, we now analyze the behavior of the scalar field in the immediate vicinity of the PBH to determine the accretion rate. In this region, the gravitational potential of the BH is the dominant factor shaping the spacetime, while the effect of the cosmic expansion is negligible. We define a characteristic transition scale, $r_{i}$, where the local gravitational energy density of the BH becomes comparable to the background energy density of the FLRW universe. Within this boundary ($r < r_i$), the dynamics are governed by a quasi-static Schwarzschild geometry, with the Hubble flow treated as a small perturbation to the accretion flow. Conversely, for $r > r_i$, the background is well-described by standard cosmological expansion parameterized by the scale factor $a(t)$. We model this geometry using the following piecewise metric~\cite{1916SPAW.......189S, Hui:2019aqm}:
\begin{equation}\label{eq:metric}
   ds^2 =  \begin{cases}
-\left(1-\frac{r_s}{r}\right)dt^2 + \left(1-\frac{r_s}{r}\right)^{-1} dr^2 + r^2 d\Omega^2 & \quad \text{for } r < r_i \\ \\
-{\cal F}_{i} dt^2 + {\cal F}_i^{-1}a(t)^2 dr^2 + a(t)^2 r^2 d\Omega^2 & \quad \text{for } r \geq r_i
\end{cases}
\end{equation}
where ${\cal F}_i \equiv (1-r_s/r_i)$ acts as a matching factor at the boundary, $r_s = 2 G M$
is the Schwarzschild radius and $d\Omega^2 = d\theta^2 + \sin^2\theta d\phi^2$ is the line element on the unit two-sphere. Here $M$ is the mass of the BH. We emphasize that this piecewise metric presented in Eq.~\eqref{eq:metric} is a phenomenological approximation. In a rigorous General Relativistic treatment, smoothly connecting a local black hole spacetime to a cosmological background requires solving the full Einstein constraint equations at the transition boundary to satisfy Hamiltonian and momentum constraints. By artificially patching the Schwarzschild and FLRW metrics at $r_i$, we are explicitly bypassing these constraint equations to maintain analytical tractability. Consequently, this patching introduces an unphysical discontinuity in the spacetime curvature at $r_i$. Our underlying physical assumption is that the cycle-averaged mass accretion rate at the event horizon is primarily dictated by the asymptotic behavior of the field in these two distinct regimes, rather than being dominated by the localized scattering effects occurring at this artificial geometric discontinuity.

The equation of motion for the scalar field in the BH spacetime is given by the Klein-Gordon equation,
\begin{equation}
\square \Phi - \frac{dV}{d\Phi} = 0,
\end{equation}
where the d'Alembertian is defined as $\square \equiv \frac{1}{\sqrt{-g}} \partial_{\mu} \left( \sqrt{-g} g^{\mu \nu} \partial_{\nu} \right)$. To solve this partial differential equation in the Schwarzschild background ($r < r_i$), we employ the method of separation of variables. We evaluate the solutions for different values of $n$. We will first solve the $n = 1$ case, followed by the $n > 1$ scenarios. Because the $n > 1$ cases render both the inflaton amplitude and the effective frequency time-dependent, careful consideration must be taken when constructing the solution ans\"{a}tze.

\noindent
We decompose the field into spherical harmonics $Y_{\ell m}(\theta,\phi)$ using the following ansatz 
\begin{equation}
\Phi(t,r,\theta,\phi) = \frac{{\cal C}^{(n)}(t)}{r} R(n,r) e^{-i\nu_n(t) t } Y_{\ell m}(\theta,\phi),
\end{equation}
where ${\cal C}^{(n)}(t)$ and $\nu_n(t)$ are slowly varying, time-dependent amplitude and frequency parameters for the first harmonics associated with the oscillating inflaton field discussed earlier, such that $|\dot{{\cal C}}^{(n)}/{\cal C}^{(n)}|\simeq 0 << |\dot{\Phi}/\Phi|$ and $\dot{\nu}_n/\nu_n \simeq 0 << |\dot{\Phi}/\Phi|$ over an oscillation cycle, and {\it those parameters we call as slow variable}. Substituting this into the equation of motion yields the radial equation  
\begin{equation}\label{eq:radial_eq}
\frac{d^2 R}{dr^{*2}} + \left[ \nu_n^2 - \left(1-\frac{r_s}{r}\right) \left( \frac{r_s}{r^3} + \frac{\ell(\ell+1)}{r^2} +  \frac{2n \Lambda^4}{M_{p}^{2n}} \left(\frac{2}{3\alpha}\right)^n e^{-i(2n-2)\nu_n  t} \left(\frac{{\cal C}^{(n)} R}{r}\right)^{2n-2} \right) \right] R = 0,
\end{equation}
where we have introduced the standard ``tortoise coordinate'' $r^*$, defined by $d r^* = (1-r_s/r)^{-1} dr$, which integrates to
\begin{equation}\label{eq:tortoise_cor}
    r^* = r + r_s \ln\left(\frac{r}{r_s}-1\right) .
\end{equation}
We will explicitly solve this radial equation for the different regimes in the upcoming sections.

\subsection{The $n = 1$ Case}
\label{subsec:n_1_case}

\begin{figure}[t]
\centering
\includegraphics[width=.51\textwidth]{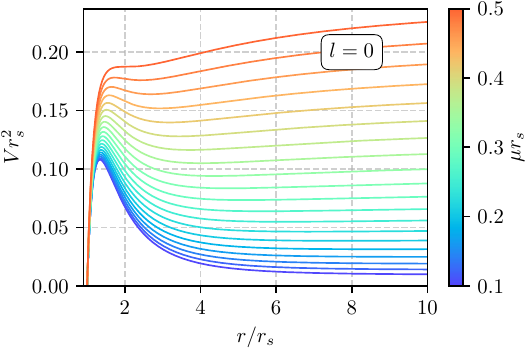}
\hspace{-0.1cm}
\includegraphics[width=.47\textwidth]{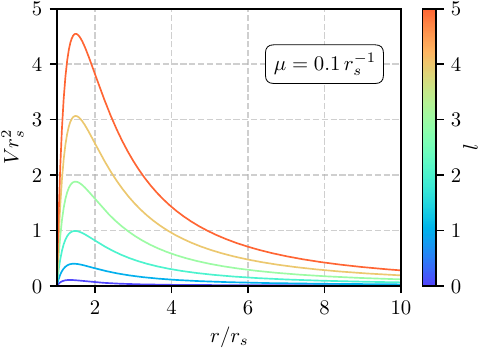}
\caption{Plots of the effective potential $V_l(r)$ from Eq.~\eqref{eq:R_eq_n_1}, shown in dimensionless units $V_l r_s^2$ as a function of $r/r_s$. \textbf{Left Panel:} The potential for the s-wave mode ($l=0$) is shown for different values of the scalar field mass, $\mu r_s$, as indicated by the color bar (ranging from 0.1 to 0.5). \textbf{Right Panel:} The potential for a fixed mass ($\mu r_s = 0.1$) is shown for different angular momentum modes, $l$ (ranging from 0 to 5). The plots illustrate that the potential barrier increases with both the field's mass and its angular momentum.}
\label{fig:effective_potential}
\end{figure}

In the $n=1$ case, the radial equation \eqref{eq:radial_eq} simplifies to
\begin{equation}
\frac{d^2 R}{dr^{*2}} + \left( \nu_1^2 - V_\ell \right) R = 0, \label{eq:R_eq_n_1}
\end{equation}
where the effective potential $V_\ell(r)$, is given by
\begin{equation}
V_\ell(r) = {\cal F}(r) \left( \frac{r_s}{r^3} +\frac{\ell(\ell+1)}{r^2} +\mu^2 \right).
\end{equation}
Here, $\mu^2 \equiv 2 \frac{\Lambda^4}{M_p^2} \left(\frac{2}{3\alpha}\right)$, with $\mu$ representing the effective mass of the inflaton field. To eliminate the first derivative term that arises when switching from tortoise to standard radial coordinates, we perform a change of variable $Z(r) = R(r) \sqrt{{\cal F}(r)}$, which yields
\begin{equation}
\frac{d^2 Z}{dr^2} + \left( \frac{r_s^2}{4r^4 {\cal F}(r)^2} + \frac{r_s}{{\cal F}(r) r^3} + \frac{\nu_1^2}{{\cal F}(r)^2} - \frac{V_\ell(r)}{{\cal F}(r)^2} \right)Z = 0.
\end{equation}
Introducing the dispersion relation $k^2 = \nu_1^2 - \mu^2$ and the shifted coordinate $x = r - r_s$, the equation can be rewritten as~\cite{Hui:2019aqm}
\begin{equation}
\frac{d^2 Z}{dx^2}
+ \left[
k^2
+ \frac{r_s^2}{4x^2(x+r_s)^2}
+ \frac{k^2 r_s}{x}
+ \frac{\nu_1^2 r_s^2}{x^2}
+ \frac{\nu_1^2 r_s}{x}
- \frac{\ell(\ell+1)}{x(x+r_s)}
\right] Z = 0.
\end{equation}
Further transforming to dimensionless quantities $\tilde{\nu_1} = \nu_1 r_s$, $\tilde{k} = k r_s$, and $y = 1 - \frac{r}{r_s}$, the equation reduces to the standard confluent Heun form:
\begin{equation}
\frac{d^2 Z}{dy^2}
+ \left[
\tilde{k}^2
- \frac{(\tilde{k}^2+\tilde{\nu_1}^2)-\ell(\ell+1)-\tfrac12}{y}
+ \frac{\tilde{\nu_1}^2+\tfrac14}{y^2}
+ \frac{1}{4(y-1)^2}
- \frac{\ell(\ell+1)+\tfrac12}{y-1}
\right] Z = 0.
\end{equation}
This differential equation admits a solution in terms of the Confluent Heun function, $\mathrm{HeunC}$, such that
\begin{equation}
Z(y) \propto e^{\frac{\alpha y}{2}} y^{\frac{1+\beta}{2}} \sqrt{1-y} \; \mathrm{HeunC}(\alpha,\beta,\gamma,\delta,\eta,y),
\end{equation}
with the coefficients given by
\begin{equation}
\alpha = \pm 2 i \tilde{k}, \quad
\beta = \pm 2 i \tilde{\nu_1}, \quad
\gamma = 0, \quad
\delta = -(\tilde{k}^2+\tilde{\nu_1}^2), \quad
\eta = \tilde{k}^2+\tilde{\nu_1}^2 - \ell(\ell+1).
\end{equation}
Returning to the radial coordinate $r$, the general solution for $R(r)$ is a linear combination of two independent branches 
\begin{equation}
\begin{split}
R(r) = & C_{\rm out} e^{i k r} \left( \frac{r}{r_s}-1 \right)^{i\nu_1 r_s} \frac{r}{r_s} \mathrm{HeunC} \!\left(
 2 i k r_s, 2 i \nu_1 r_s, 0, \delta, \eta, 1-\frac{r}{r_s} \right) \\
& \quad  + C_{\rm in} e^{-i k r} \left( \frac{r}{r_s}-1 \right)^{-i\nu_1 r_s} \frac{r}{r_s} \mathrm{HeunC} \!\left(
 -2 i k r_s, -2 i \nu_1 r_s, 0, \delta, \eta, 1-\frac{r}{r_s} \right),
\end{split}
\end{equation}
where $C_{\rm out}$ and $C_{\rm in}$ are integration constants. For the purpose of evaluating accretion, the spherically symmetric $s$-wave mode ($\ell=0$) provides the dominant contribution. Therefore, the general solution for the $s$-wave field $\Phi(t,r)$ is a superposition of ingoing and outgoing waves 
\begin{equation}
\begin{split}
& \Phi(t,r)  = \frac{\mathcal{C}^{(1)}_{\rm out}}{r_s} e^{-i\nu_1 t} e^{i k r} \left( \frac{r}{r_s}-1 \right)^{i\nu_1 r_s} \mathrm{HeunC} \!\left(
 2 i k r_s, 2 i \nu_1 r_s, 0, -k^2 r_s^2-\nu_1^2 r_s^2, k^2 r_s^2+\nu_1^2 r_s^2, 1-\frac{r}{r_s} \right) \\
&  + \frac{\mathcal{C}^{(1)}_{\rm in}}{r_s} e^{-i\nu_1 t} e^{-i k r} \left( \frac{r}{r_s}-1 \right)^{-i\nu_1 r_s} \mathrm{HeunC} \!\left(
 -2 i k r_s, -2 i \nu_1 r_s, 0, -k^2 r_s^2-\nu_1^2 r_s^2, k^2 r_s^2+\nu_1^2 r_s^2, 1-\frac{r}{r_s} \right).
\end{split}
\label{eq:general_phi_sol}
\end{equation}
To understand the physical behavior of this solution, we examine its form in two important limits. First, in the near-horizon limit ($r \to r_s$), the $\mathrm{HeunC}$ function approaches unity. Enforcing the physical boundary condition of purely ingoing waves at the horizon (setting $\mathcal{C}^{(1)}_{\rm out} = 0$), the solution simplifies to
\begin{equation}
\Phi (t,r) \simeq \frac{\mathcal{C}^{(1)}_{\rm in}}{r_s} e^{-i\nu_1 t} e^{-i k r} \left( \frac{r}{r_s}-1 \right)^{-i\nu_1 r_s} \simeq \frac{\mathcal{C}^{(1)}_{\rm in}}{r_s} e^{-i\nu_1( t + r^*)},
\end{equation}
where the constant phase shifts have been absorbed into amplitude $\mathcal{C}^{(1)}_{\rm in}$. This is manifestly a purely ingoing wave. Assuming this constant is real and carries the time dependence inherited from the cosmological background, the real part of the scalar field is given by
\begin{equation}
    \Phi(t,r) \simeq \frac{\mathcal{C}^{(1)}_{\rm in}(t)}{r_s}\cos(\nu_1 (t+r^*)).
\label{eq:near_horizon_sol}
\end{equation}
Second, to facilitate matching to the cosmological solution at the boundary $r \sim r_i$, we require the asymptotic behavior of the radial solution in the intermediate region $r_s \ll r \lesssim r_i$. We evaluate this under the parametric assumptions~\cite{Hui:2019aqm}
\begin{equation}
\mu r_s \ll 1, \qquad \frac{r}{r_s} \gg 1, \quad \text{but} \quad \mu^2 r_s^2 \frac{r}{r_s} \lesssim 1.
\end{equation}
These conditions define the transition radius as $r_i = 1/\mu^2 r_s$. We restrict our attention to the zero-momentum mode, $k=0$ (where $\nu_1 = \mu$), for which the relevant branch of the scalar field reduces to
\begin{equation}
\Phi(t,r) = \frac{\mathcal{C}^{(1)}_{\rm in}}{r_s} e^{-i\mu t} \left( \frac{r}{r_s}-1 \right)^{-i\mu r_s} \mathrm{HeunC} \!\left(
0, \pm 2 i \mu r_s, 0, -\mu^2 r_s^2, \mu^2 r_s^2, 1-\frac{r}{r_s} \right).
\end{equation}
Using the small $\mu r_s$ expansion of the confluent Heun function (see Appendix \ref{app:heun_function}) and keeping terms up to $\mathcal{O}(\mu^2 r r_s)$, we obtain
\begin{equation}
\Phi(t,r) \simeq \frac{\mathcal{C}^{(1)}_{\rm in}}{r_s} e^{-i\mu t} \left[ 1 - \frac{1}{2}\mu^2 r r_s - i \mu r_s \ln\!\left(1-\frac{r_s}{r}\right) \right].
\end{equation}
In the large $r$ limit ($r \gg r_s$), the logarithm admits the expansion
\begin{equation}
\ln\!\left(1-\frac{r_s}{r}\right) = - \frac{r_s}{r} - \frac{r_s^2}{2 r^2} + \mathcal{O}\!\left(\frac{1}{r^3}\right),
\end{equation}
so that the field becomes
\begin{equation}
\Phi(t,r) \simeq \frac{\mathcal{C}^{(1)}_{\rm in}}{r_s} e^{-i\mu t} \left[ 1 - \frac{1}{2}\mu^2 r r_s + i \mu r_s \frac{r_s}{r} + \mathcal{O}\!\left(\frac{1}{r^2}\right) \right].
\end{equation}
If the hierarchy $\mu r_s \ll (r_s/r_i)^2$ is satisfied, the term proportional to $\mu^2 r r_s$ is subdominant in the matching region, and the expression simplifies to
\begin{equation}
\Phi(t,r) \simeq \frac{\mathcal{C}^{(1)}_{\rm in}}{r_s} e^{-i\mu t} \left( 1 + i \mu \frac{r_s^2}{r} \right).
\end{equation}
Assuming the field is real, we take the real part of this expression. Allowing the amplitude $\mathcal{C}^{(1)}_{\rm in}$ to act as an envelope carrying the slow time evolution from the cosmological background, the real scalar field profile is approximated by
\begin{equation}
    \Phi(t,r) \simeq \frac{\mathcal{C}^{(1)}_{\rm in}(t)}{r_s} \left[ \cos{(\mu t)} - \frac{\mu r_s^2}{r}\sin{(\mu t)} \right] \approx \frac{\mathcal{C}^{(1)}_{\rm in}(t)}{r_s} \cos{(\mu t)} ,
\label{eq:matching_solution}
\end{equation}
The second approximation holds provided that the correction term remains subdominant, $\mu^2 r_s^2 \frac{r}{r_s} \lesssim 1 ,$ so that the $r^{-1}$ contribution can be safely neglected. This expression~\eqref{eq:matching_solution} dictates the form of the scalar field within the gravitational influence of the BH and serves as the crucial link for matching with the cosmological FLRW solution at the boundary $r_i$.

It is crucial to clarify the physical nature of the boundary conditions applied to the scalar field. During the reheating epoch, the inflaton field in the far-zone FLRW background behaves as a coherently oscillating, spatially homogeneous condensate, rather than a flux of traveling waves propagating from infinity. Therefore, at the matching radius $r_i$, our boundary condition requires the local scalar field to smoothly match this spatially homogeneous, oscillating cosmological background. Conversely, at the BH event horizon, we impose a purely ingoing boundary condition. By solving the exact Klein-Gordon wave equation between these boundaries, our solution naturally accounts for the reflection and scattering of the scalar field off the BH's gravitational potential barrier. This wave-scattering effect significantly restricts the ingoing flux, naturally preventing overestimation of the accretion rate.

\subsection{The $n > 1$ Case}
\label{subsec:n_gt_1_case}
For $n>1$, the nonlinear term in the potential induces an explicit time dependence in the radial equation \eqref{eq:radial_eq}. To obtain a tractable analytic equation, we perform a time average over one complete oscillation period as
\begin{equation}
\frac{d^2 R}{dr^{*2}} + \left[ \left\langle \nu_n^2 \right\rangle - {\cal F}(r) \left( \frac{r_s}{r^3} + \frac{\ell(\ell+1)}{r^2} +  \frac{2n \Lambda^4}{M_{p}^{2n}} \left(\frac{2}{3\alpha}\right)^n \left\langle e^{-i(2n-2)\nu_n  t} \right\rangle \left(\frac{{\cal C}^{(n)} R}{r}\right)^{2n-2} \right) \right] R = 0,
\end{equation}
Assuming the frequency $\nu_n(t)$ evolves slowly compared to the oscillation timescale, the highly oscillatory contribution $\propto e^{-i(2n-2)\nu_n t}$ averages to zero. The radial equation then reduces to a standard Schr\"{o}dinger-like form 
\begin{equation}\label{eq:radial_n}
\frac{d^2 R}{dr^{*2}} + \left( \nu_n^2 - V_\ell(r) \right) R = 0,
\end{equation}
where the effective potential is given by
\begin{equation}\label{eq:potential_eff_n}
V_\ell(r) = {\cal F}(r) \left( \frac{r_s}{r^3} + \frac{\ell(\ell+1)}{r^2} \right).
\end{equation}
In the near-horizon limit ($r \to r_s$), the metric factor ${\cal F}(r) \to 0$, causing the effective potential to vanish ($V_\ell(r) \to 0$). The radial equation simplifies to
\begin{equation}
\frac{d^2 R}{dr^{*2}} + \nu_n^2 R = 0,
\end{equation}
which yields standard plane-wave solutions in the tortoise coordinate, $R(r^*) \propto e^{\pm i \nu_n r^*}$. Imposing the physical boundary condition of purely ingoing waves at the horizon, we select the positive sign. The scalar field in this regime therefore behaves as
\begin{equation}
\Phi(t,r) \simeq \frac{\mathcal{C}^{(n)}_{\rm in}}{r_s} e^{-i\nu_n (t + r^*)},
\end{equation}
where $\mathcal{C}^{(n)}_{\rm in}$ is the near-horizon amplitude for the ingoing mode. Therefore the real part of the scalar field is given by
\begin{equation}\label{eq:near_horizon_sol_n}
    \Phi(t,r) \simeq \frac{\mathcal{C}^{(n)}_{\rm in}(t)}{r_s}\cos(\nu_n (t+r^*)).
\end{equation}
We now consider the asymptotic intermediate regime $r \gg r_s$. Motivated by the necessity of cosmological matching, we seek solutions that become effectively spatially homogeneous at large distances from the BH ($r_i \gg r_s$). In this limit, the tortoise coordinate~\eqref{eq:tortoise_cor} is well-approximated by the standard radial coordinate ($r^* \simeq r$), and the horizon function approaches unity (${\cal F}(r) \equiv 1 - r_s/r \simeq 1$). Restricting our analysis to the dominant $s$-wave ($\ell=0$) mode, the effective potential~\eqref{eq:potential_eff_n} simplifies to $ V_0(r) \simeq  r_s/r^3 $. The behavior of the radial equation is governed by the competition between the temporal frequency term $\nu_n^2$ and the spatial effective potential $V_0(r)$ (see Eq.~\ref{eq:radial_n}). To capture the intermediate region where the spatial curvature of the field is dictated primarily by the BH's gravitational pull rather than the field's own temporal oscillation, we examine the low-frequency limit where the potential dominates such that $\nu_n^2 \lesssim r_s/r^3$. This yields the transition radius $r_i = (r_s/\nu_n^2)^{1/3}$. The radial equation then simplifies to
\begin{equation}
\frac{d^2 R}{dr^2} - \frac{r_s}{r^3} R = 0.
\end{equation}
This differential equation admits an analytic solution in terms of modified Bessel functions 
\begin{equation}
R(r) = C_1 \sqrt{\frac{r}{r_s}} \, I_1\!\left( 2 \sqrt{\frac{r_s}{r}} \right) + 2 C_2 \sqrt{\frac{r}{r_s}} \, K_1\!\left( 2 \sqrt{\frac{r_s}{r}} \right),
\end{equation}
where $I_\alpha(x)$ and $K_\alpha(x)$ denote the modified Bessel functions of the first and second kind, respectively. For small arguments ($|x| \ll 1$, corresponding to large $r$), their leading-order expansions are
\begin{equation}
\begin{split}
I_\alpha(x) &\sim \frac{1}{\Gamma(\alpha+1)} \left( \frac{x}{2} \right)^\alpha, \\
K_\alpha(x) &\sim \frac{\Gamma(\alpha)}{2} \left( \frac{2}{x} \right)^\alpha \qquad (\alpha > 0).
\end{split}
\end{equation}
Applying these limits of small-arguments to our solution, we find that $R(r) \sim C_1 + C_2 (r/r_s)$. Reconstructing the full scalar field $\Phi(t,r)$, we obtain the asymptotic behavior 
\begin{equation}
\Phi(t,r) \simeq \frac{\mathcal{C}^{(n)}_{\rm in}}{r_s} e^{-i\nu_n t} \left( 1 + C_3 \frac{r_s}{r} \right), \qquad \text{for} \quad r_s \ll r \lesssim \left( \frac{r_s}{\nu_n^2} \right)^{1/3}.
\end{equation}
As long as $r_s/r \ll 1$, the correction $1/r$ is subdominant and the field profile is primarily spatially homogeneous.
\begin{equation}
\Phi(t,r) \simeq \frac{\mathcal{C}^{(n)}_{\rm in}}{r_s} e^{-i\nu_n t},
\end{equation}
where the normalization constant $\mathcal{C}^{(n)}_{\rm in}$ dictates the amplitude and will be fixed by continuously matching to the cosmological FLRW background at the boundary $r_i$. Therefore, the real part of the scalar field is given by
\begin{equation}
    \Phi(t,r) \simeq \frac{\mathcal{C}^{(n)}_{\rm in}(t)}{r_s}\cos(\nu_n t).
\label{eq:matching_solution_n}
\end{equation}
With the scalar field behavior fully determined in both the cosmological far-zone and the BH near-zone, we now proceed to match these solutions at the transition boundary to derive the mass accretion rate.


\section{Asymptotic Matching and the Mass Accretion Rate}
\label{sec:matching_accretion}

The final step in our theoretical framework is to connect the near-horizon dynamics of the scalar field to its cosmological evolution. The near-horizon solution depends on a slowly varying, time-dependent amplitude, $\mathcal{C}^{(n)}_{\rm in}(t)$, while the far-field solution describes the full evolution of the inflaton envelope driven by cosmic expansion. The transition radius $r = r_i$ is physically fixed by the scale at which the effective frequency of the inner zone matches the characteristic oscillation frequency of the cosmological far-zone. To quantitatively justify the spatial hierarchy assumed during the matching procedure, we examine the temporal evolution of the transition radius $r_i$. For a quadratic potential ($n=1$), the effective mass of the scalar field ($\mu$) is strictly constant, leading to a static transition boundary located at roughly $r_i \simeq 2 r_s$. Conversely, for higher-order potentials ($n>1$), the effective frequency (see Eq.~\ref{eq:omega_n}) of the inflaton field explicitly depends on its amplitude, which decreases as the universe expands due to cosmic dilution ($\nu_n \propto t^{-(n-1)/n}$). Because the far-zone frequency drops over time, the matching boundary must migrate outward to compensate. As the universe expands, this ratio grows rapidly, easily reaching $r_i \gg r_s$ at later times. This evolutionary behavior strictly validates our earlier asymptotic assumption that the intermediate matching region lies well outside the immediate gravitational well of the BH for the non-linear potential cases. By enforcing continuity of the field amplitudes exactly at this specific matching boundary, we uniquely determine $\mathcal{C}^{(n)}_{\rm in}(t)$ and thus obtain the complete form of the scalar field driving the accretion.

\subsection{Amplitude Matching at the Boundary}

As established in Sections \ref{sec:bh_spacetime} and \ref{sec:flrw_background}, the real part of the scalar field in the intermediate matching region ($r_s \ll r \lesssim r_i$) is spatially homogeneous. The solutions in this regime take the form 
\begin{equation}
\Phi (t,r) \simeq 
\begin{cases}
\mathcal{C}^{(n)}_{\rm in}(t) \cos(\nu_n t ) \quad & \text{(Near-zone extrapolated outward)} \\
\mathcal{P}_1^{(n)} \Phi_{\rm end} \left( \frac{t_{\rm end}}{t} \right)^{1/n} \cos(\nu_n t) \quad & \text{(Far-zone extrapolated inward)}
\end{cases}
\end{equation}
where we have absorbed the geometric factor $r_s$ from Eqs.~\eqref{eq:matching_solution} and \eqref{eq:matching_solution_n} into the effective amplitude $\mathcal{C}^{(n)}_{\rm in}(t)$. The expression for $\nu_n$ is given by Eq.~\eqref{eq:omega_n}. Equating these expressions yields the time evolution of the near-horizon amplitude 
\begin{equation}
\mathcal{C}^{(n)}_{\rm in}(t) = \mathcal{P}_1^{(n)} \Phi_{\rm end} \left( \frac{t_{\rm end}}{t} \right)^{1/n},
\label{eq:C_n_matched}
\end{equation}
where $\mathcal{P}_1^{(1)} = 1$, $\mathcal{P}_1^{(2)} \approx 0.47$, and $\mathcal{P}_1^{(3)} \approx 0.46$ characterize the fundamental harmonic mode for each potential shape. Substituting this back into the purely ingoing near-horizon solution (Eq.~\eqref{eq:near_horizon_sol} and \eqref{eq:near_horizon_sol_n}), the field profile evaluating at the event horizon ($r \to r_s$) becomes
\begin{equation}
\Phi_n(t,r) \simeq \mathcal{C}^{(n)}_{\rm in}(t) \cos(\omega_n(t+r^*)).
\label{eq:full_matched_field}
\end{equation}

\subsection{Calculation of the Accretion Rate}

Once the full near-horizon solution is known, we compute the rate of mass accretion onto the PBH. The accretion rate is determined by the integrated flux of energy flowing across the event horizon. This is governed by the mixed time-radial component of the scalar field's energy-momentum tensor, $T^r_t = g_{t\mu}T^{r\mu} = -f(r) T^{rt}$, where $T^{rt} = -(\partial_t\Phi)(\partial_r\Phi)$. The total mass accretion rate is then~\cite{Hui:2019aqm, Michel1972ApSS, Harada:2004pf, Urena-Lopez:2011lcm}
\begin{equation}
    \frac{dM}{dt} = \oint_{r=r_s} \sqrt{-g} \, T^r_t \, d\theta d\phi = 4\pi r_s^2 \, T^r_t \big|_{r=r_s}.
\end{equation}
Using Eq.~\eqref{eq:full_matched_field} and taking the derivatives $\partial_r$ and $\partial_t$ (noting that $\dot{\nu}_n \simeq 0$ and $\dot{\mathcal{C}}^{(n)}_{\rm in} \simeq 0$ in the adiabatic limit), the relevant energy-momentum component is 
\begin{equation}
T^r_t \simeq {\mathcal{C}^{(n)}_{\rm in}}^2 \nu_n^2 \sin^2(\omega_n(t+r^*))  .
\end{equation}
Because the accretion timescale is much longer than the oscillation period, we time-average over one cycle, gives
\begin{equation}
\langle T^r_t \rangle \simeq \frac{1}{2} {\mathcal{C}^{(n)}_{\rm in}}^2 \nu_n^2.
\end{equation}
Therefore, the cycle-averaged mass accretion rate at the horizon reduces to a remarkably simple form 
\begin{equation}
\dot{M} \simeq 2 \pi r_s^2 {\mathcal{C}^{(n)}_{\rm in}}^2 \nu_n^2.
\label{eq:M_dot_general}
\end{equation}
To integrate Eq.~\eqref{eq:M_dot_general}, we must evaluate the time dependence of the product $\mathcal{C}^{(n)}_{\rm in}(t) \nu_n(t)$. The inflaton field mode frequency $\nu_n$ scales with the potential curvature, $\nu_n \propto \Phi_0^{n-1}$. Using the matched amplitude from Eq.~\eqref{eq:C_n_matched} and the cosmological scaling $\Phi_0 \propto t^{-1/n}$, a remarkable cancellation occurs. All intermediate time dependencies strictly collate such that the product scales inversely with time 
\begin{equation}
\mathcal{C}^{(n)}_{\rm in}(t) \nu_n(t) = \frac{\delta_n}{t},
\end{equation}
where $\delta_n$ is a time-independent constant grouping the inflationary and potential parameters. Following algebraic simplification using the Friedmann equations at the end of inflation, this constant evaluates exclusively to 
\begin{equation}
\delta_n = \frac{2}{3} \sqrt{\pi} (n+1) \frac{\Gamma\left(\frac{1}{2n} + \frac{1}{2}\right)}{\Gamma\left(\frac{1}{2n}\right)} \mathcal{P}_1^{(n)} M_p.
\end{equation}
Substituting this into Eq.~\eqref{eq:M_dot_general}, and expressing the Schwarzschild radius as $r_s = 2GM = M / (4\pi M_p^2)$, the accretion differential equation becomes 
\begin{equation}
\dot{M} = \frac{\delta_n^2}{8\pi M_p^4} \frac{M^2}{t^2}.
\end{equation}
This indicates that the accretion rate is proportional to the square of the PBH mass and falls off as $t^{-2}$ strictly due to the background cosmological expansion, regardless of the potential power $n$.

\begin{figure}[t]
\centering

\begin{subfigure}{0.495\textwidth}
    \centering
    \includegraphics[width=\linewidth]{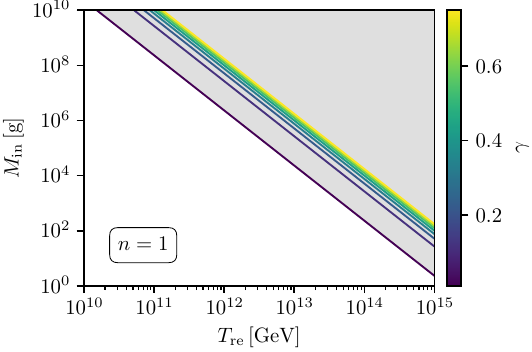}
    \caption{}
\end{subfigure}
\hfill
\begin{subfigure}{0.495\textwidth}
    \centering
    \includegraphics[width=\linewidth]{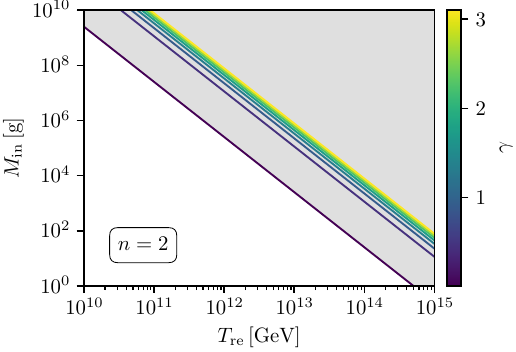}
    \caption{}
\end{subfigure}

\begin{subfigure}{0.51\textwidth}
    \centering
    \includegraphics[width=\linewidth]{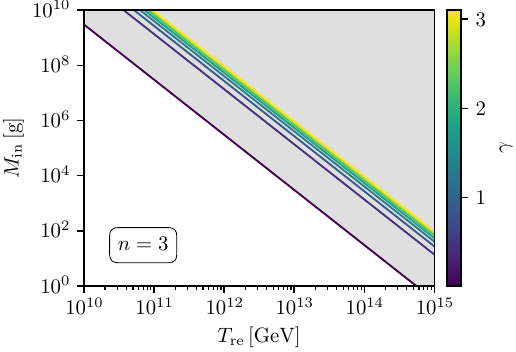}
    \caption{}
\end{subfigure}

\caption{Parameter space in the $M_{\rm in}$--$T_{\rm re}$ plane illustrating the boundary where radiation accretion overtakes inflaton accretion at the end of the reheating epoch. The three panels correspond to the background inflaton potentials (a) $n=1$, (b) $n=2$, and (c) $n=3$. The plots are generated using the $\alpha$-attractor parameter $\alpha = 1$. For $n=1$ and $n=3$, fixing $T_{\rm re}$ uniquely determines the CMB spectral index $n_s$. Because of the expansion history degeneracy for the $n=2$ case, a benchmark value of $n_s = 0.965$ is explicitly chosen to fix the inflationary boundary conditions. The colored solid lines represent the exact equality of Eq.~\eqref{eq:Tre_min_inequality} evaluated for various values of the initial mass fraction $\gamma$. Within the gray shaded regions (above the contours), the accretion rate from the relativistic thermal bath exceeds that of the oscillating scalar field.}
\label{fig:rad_domination}
\end{figure}

\section{Accretion of Both Inflaton and Radiation}
\label{sec:radiation_accretion}

Because we are analyzing PBHs during the reheating era, we must account for the evolving composition of the universe. As the inflaton decays, the surrounding environment becomes a mixture of the oscillating inflaton field and a growing relativistic thermal bath. For PBHs formed near the end of reheating, or actively accreting during this transition, radiation accretion cannot be neglected. The total mass accretion rate is the sum of the contributions from both fluids
\begin{equation}
\dot{M}_{\rm total} = \dot{M}_\Phi + \dot{M}_r.
\end{equation}
The radiation accretion rate in the unperturbed FLRW background is given by~\cite{Das:2025vts, Kalita:2025fcs}
\begin{equation}
\dot{M}_r = \frac{\lambda_c}{16 \pi M_p^4} M^2 \rho_r,
\end{equation}
where $\lambda_c \simeq 10.4$ is the dimensionless accretion efficiency for a radiation fluid, and $\rho_r$ is the background radiation energy density whose evolution is governed by the reheating dynamics derived in Section \ref{sec:reheating_dynamics}.

\subsection{Accretion During Reheating ($a < a_{\rm re}$)}


Prior to the completion of reheating, the PBH accretes from both the dominant inflaton field and the subdominant, but growing, radiation bath. Expressing both accretion rates in terms of the scale factor $a$ and utilizing the relation $\rho_r \propto a^{-\frac{3}{2}(1+\omega_\Phi)}$, the combined accretion differential equation becomes
\begin{equation}
H \left( \frac{a}{a_{\rm in}} \right) \frac{dM}{d\left( \frac{a}{a_{\rm in}} \right)} = C_\Phi M^2 \left( \frac{a}{a_{\rm in} } \right)^{-3(1+\omega_\Phi)} + C_r M^2 \left( \frac{a}{a_{\rm in} } \right)^{-\frac{3}{2}(1+\omega_\Phi)},
\end{equation}
where $C_\Phi = \delta_n^2 / (8\pi M_p^4 t_{\rm in}^2)$ and $C_r = \lambda_c \rho_r^{\rm in} / (16\pi M_p^4)$ define the accretion efficiencies of the two components. Integrating this equation from the formation scale factor $a_{\rm in}$ to an arbitrary scale factor $a < a_{\rm re}$ yields the modified PBH mass evolution
\begin{equation}\label{eq:mass_evolution_acc}
M(a) \simeq M_{\rm in} \left\{ 1 - \frac{\delta_n^2 \gamma}{2M_p^2} \frac{3n}{n+1} \left[ 1 - \left( \frac{a}{a_{\rm in}} \right)^{-\frac{3}{2}(1+\omega_\Phi)} \right] - \frac{\lambda_c}{8\pi} \frac{3(1+\omega_\Phi)}{(5-3\omega_\Phi)} \frac{\Gamma_\Phi M_{\rm in}}{M_p^2} \ln\left( \frac{a}{a_{\rm in}} \right) \right\}^{-1}.
\end{equation}
The logarithmic term represents the continuous mass contribution from the decaying inflaton's radiation products. By comparing the two subtractive terms in the denominator, we can extract the strict physical condition determining when radiation accretion dominates over inflaton accretion. Evaluating this at the end of reheating ($a = a_{\rm re}$), radiation accretion becomes dominant if the reheating temperature and initial PBH mass satisfy
\begin{equation}\label{eq:Tre_min_inequality}
\begin{split}
\left( \frac{T_{\rm re}}{M_p} \right)^2 \left( \frac{M_{\rm in}}{M_p} \right) &\gtrsim \frac{4\sqrt{3}\pi \gamma }{\lambda_c} \left( \frac{2\sqrt{\pi}}{3} (n+1) \frac{\Gamma\left(\frac{1}{2n}+\frac{1}{2}\right)}{\Gamma\left(\frac{1}{2n}\right)} \mathcal{P}_1^{(n)} \right)^2   (1+\omega_\Phi) \frac{\left\{ 1 - \left( \frac{a_{\rm re}}{a_{\rm in}} \right)^{-\frac{3}{2}(1+\omega_\Phi)} \right\} }{\ln\left( \frac{a_{\rm re}}{a_{\rm in}} \right)}.
\end{split}
\end{equation}
This inequality demonstrates that for sufficiently high reheating temperatures or massive initial PBHs, the radiation bath plays a non-trivial role in the final asymptotic mass, counteracting the rapid dilution of the scalar field. To build physical intuition for this threshold, we note that the parameter dependence is dominated by the scaling $M_{\rm in} \propto T_{\rm re}^{-2}$. For initial mass fractions near their critical limits ($\gamma \to \gamma_c$), the right-hand side evaluates to an $\mathcal{O}(10^{-3})$ to $\mathcal{O}(1)$ geometrical factor depending on the potential index $n$. Consequently, for a typical intermediate reheating temperature of $T_{\rm re} \sim 10^{10} \text{ GeV}$, this inequality dictates that PBHs with initial masses $M_{\rm in} \gtrsim 10^8 \text{ g}$ will experience dominant radiation accretion before the epoch concludes, firmly placing this mechanism within phenomenologically relevant mass ranges.

To visualize the physical regime where the accretion of the newly formed radiation bath becomes the primary mechanism for PBH mass growth prior to the completion of reheating, we evaluate the threshold condition derived in Eq.~\eqref{eq:Tre_min_inequality}. In Fig.~\ref{fig:rad_domination}, we map this parameter space by plotting the initial PBH mass $M_{\rm in}$ against the reheating temperature $T_{\rm re}$ for the three potential shapes.

Because the dominant terms in the inequality scale approximately as $M_{\rm in} \propto T_{\rm re}^{-2}$, the boundary separating the two accretion regimes forms a distinct linear threshold with a slope of $-2$ on a logarithmic scale. The gray shaded regions (lying above the threshold lines) dictate the parameter space where radiation accretion strictly overtakes inflaton accretion by the time the universe reaches $a = a_{\rm re}$. Furthermore, as the initial mass fraction $\gamma$ increases (indicated by the colored contour lines), the boundary shifts slightly upward. This implies that for PBHs forming with a larger initial fraction of the horizon mass, a correspondingly higher initial mass or a hotter reheating temperature is required for the radiation fluid to dominate the accretion process.

\subsection{Accretion in the Radiation-Dominated Era ($a > a_{\rm re}$)}


Once the scale factor exceeds $a_{\rm re}$, the universe becomes fully radiation-dominated ($\omega_r = 1/3$). The inflaton field is entirely depleted, and the PBH relies solely on radiation accretion. The differential equation simplifies to
\begin{equation}
\frac{dM}{d(a/a_{\rm re})} = \frac{\lambda_c \rho_{\rm re}}{16\pi M_p^4 H_{\rm re}} M^2 \left( \frac{a}{a_{\rm re}} \right)^{-3},
\end{equation}
where the Hubble parameter at the onset of radiation domination is $H_{\rm re} = \sqrt{2\rho_{\rm re} / 3M_p^2}$. Integrating this from the end of reheating, where the mass is $M_{\rm re} \equiv M(a_{\rm re})$, to an arbitrary scale factor $a$, gives the late-time mass evolution
\begin{equation}
M(a) \simeq M_{\rm re} \left\{ 1 - \frac{\sqrt{3} \lambda_c}{32\sqrt{2}} \left( \frac{g_*}{30} \right)^{\frac{1}{2}} \left( \frac{M_{\rm re}}{M_p} \right) \left( \frac{T_{\rm re}}{M_p} \right)^2 \left[ 1 -  \left( \frac{a_{\rm re}}{a_{\rm in}} \right)^{2} \left( \frac{a}{a_{\rm in}} \right)^{-2} \right] \right\}^{-1}.
\end{equation}
Because the radiation energy density dilutes rapidly as $a^{-4}$ in this epoch, the accretion rate drops off sharply. As $a \gg a_{\rm re}$, the mass quickly converges to a final, stable limit, finalizing the accretion phase of the PBH's lifecycle.

\subsection{Asymptotic Mass Limit and Runaway Accretion}
\label{sec:M_acc}


\begin{figure}[t]
\centering
\includegraphics[width=0.55\linewidth]{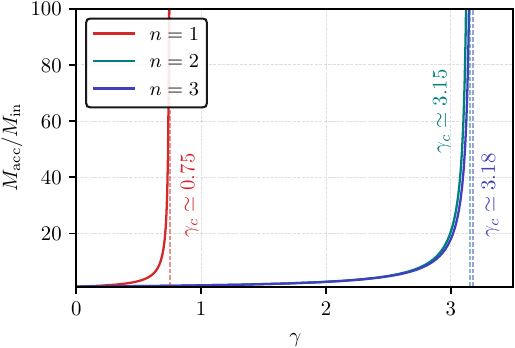} 
\caption{The asymptotic accreted mass $M_{\rm acc}/M_{\rm in}$ as a function of the initial mass fraction $\gamma$. The mass strictly diverges as $\gamma$ approaches a potential-dependent critical value $\gamma_c$. The runaway accretion limits are numerically evaluated as $\gamma_c \simeq 0.75$ for $n=1$, $\gamma_c \simeq 3.15$ for $n=2$, and $\gamma_c \simeq 3.18$ for $n=3$.}
\label{fig:gamma_divergence}
\end{figure}

From Eq.~\eqref{eq:mass_evolution_acc}, we can approximate the maximum mass a PBH can achieve through accretion. As the universe expands far beyond the PBH formation epoch ($a \gg a_{\rm in}$), the expansion term $(a/a_{\rm in})^{-3(1+\omega_\Phi)/2}$ decays to zero. By neglecting the late-time enhancement from the radiation component, we obtain the net asymptotic PBH mass, $M_{\rm acc}$, as
\begin{equation}
M_{\rm acc} \approx M_{\rm in} \left\{ 1 - \frac{2\pi}{3} \gamma \, n(n+1) \left( \mathcal{P}_1^{(n)} \frac{\Gamma\left(\frac{1}{2n} + \frac{1}{2}\right)}{\Gamma\left(\frac{1}{2n}\right)} \right)^2 \right\}^{-1}.
\label{eq:M_acc}
\end{equation}
Crucially, Eq.~\eqref{eq:M_acc} reveals a strict physical limit on the initial conditions. If the initial mass fraction $\gamma$ is sufficiently large, the denominator in Eq.~\eqref{eq:M_acc} approaches zero, causing the accreted mass to diverge. This critical initial mass fraction, $\gamma_c$, is found by setting the term inside the braces to zero
\begin{equation}
\gamma_c \approx \frac{3}{2\pi} \frac{1}{n(n+1)} \left( \mathcal{P}_1^{(n)} \frac{\Gamma\left(\frac{1}{2n} + \frac{1}{2}\right)}{\Gamma\left(\frac{1}{2n}\right)} \right)^{-2}.
\label{eq:gamma_critical}
\end{equation}
Physically, when a PBH forms with an initial mass fraction approaching the critical value, $\gamma \rightarrow \gamma_c$, its accretion rate outpaces the background cosmological dilution. In this regime, the PBH undergoes runaway accretion, potentially consuming the entire horizon mass.

We plot the dependence of the asymptotic mass $M_{\rm acc}/M_{\rm in}$ on $\gamma$ in Fig.~\ref{fig:gamma_divergence}. The divergence is distinctly visible at the critical thresholds: $\gamma_c \simeq 0.75$ for the quadratic potential ($n=1$), $\gamma_c \simeq 3.15$ for the quartic potential ($n=2$), and $\gamma_c \simeq 3.18$ for the sextic potential ($n=3$). For physically realistic PBH populations that do not overclose the universe or consume the entire reheating bath, formation mechanisms must strictly satisfy $\gamma < \gamma_c$.

\begin{figure}[t]
\centering

\begin{subfigure}{0.49\textwidth}
    \centering
    \includegraphics[width=\linewidth]{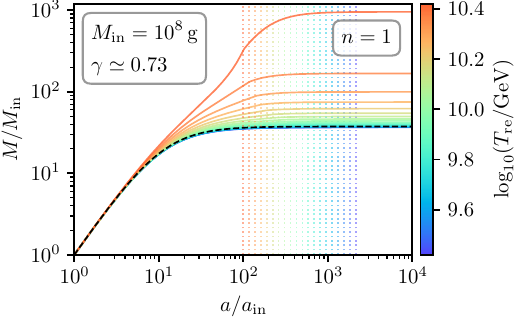}
    \caption{}
\end{subfigure}
\hfill
\begin{subfigure}{0.50\textwidth}
    \centering
    \includegraphics[width=\linewidth]{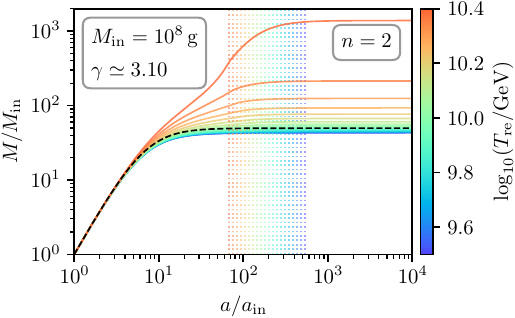}
    \caption{}
\end{subfigure}


\begin{subfigure}{0.51\textwidth}
    \centering
    \includegraphics[width=\linewidth]{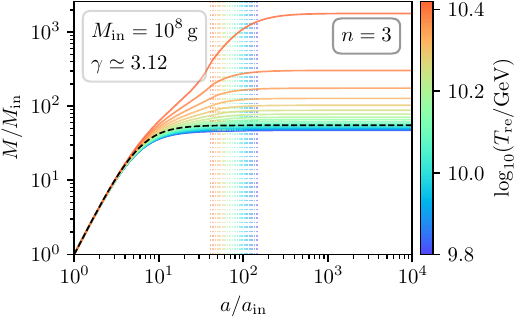}
    \caption{}
\end{subfigure}

\caption{Evolution of the PBH mass ratio $M/M_{\rm in}$ highlighting the contribution of radiation accretion during the reheating epoch. The panels correspond to inflaton potentials with (a) $n=1$, (b) $n=2$, and (c) $n=3$, evaluated at initial mass fractions closely approaching their critical runaway limits ($\gamma \simeq 0.73$, $3.10$, and $3.12$, respectively) for an initial mass of $10^8 \, \rm g$. The solid curves trace the full mass evolution including both inflaton and radiation accretion, with the color map denoting the reheating temperature $\log_{10}(T_{\rm re}/\text{GeV})$. The black dashed curve isolates the baseline evolution assuming solely inflaton accretion. Vertical dashed lines mark the end of reheating ($a = a_{\rm re}$) for each corresponding $T_{\rm re}$, after which the universe becomes radiation-dominated and the mass growth rapidly plateaus. The range of $T_{\rm re}$ is selected to illustrate the specific regime where the relativistic thermal bath significantly enhances the final accreted mass.}
\label{fig:radiation_accretion_impact}
\end{figure}

With the cosmological observables explicitly linked to the reheating dynamics, we can evaluate the full PBH accretion history using physically motivated boundary conditions. To compute the mass evolution, we must fix the underlying parameters of the inflationary potential and the thermal history. For our numerical evaluation, we set the $\alpha$-attractor scale parameter to $\alpha = 1$.
As demonstrated in Section \ref{sec:cmb_matching}, specifying $T_{\rm re}$ uniquely determines the duration of reheating ($N_{\rm re}$) and subsequently fixes the scalar spectral index $n_s$ for the $n=1$ and $n=3$ potentials. However, the $n=2$ case presents a unique degeneracy: because the $\Phi^4$ inflaton energy density dilutes at the exact same rate as radiation ($a^{-4}$), the expansion history is blind to the transition. Consequently, $T_{\rm re}$ is completely decoupled from $n_s$. To evaluate the initial conditions at the end of inflation ($\Phi_{\rm end}$ and $\rho_{\rm end}$) for the $n=2$ scenario, we adopt a benchmark value of $n_s = 0.965$.


\begin{figure}[t]
\centering

\begin{subfigure}{0.495\textwidth}
    \centering
    \includegraphics[width=\linewidth]{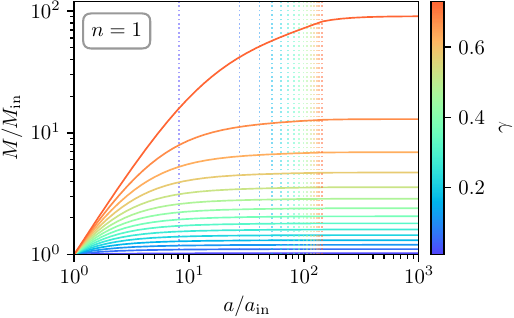}
    \caption{}
\end{subfigure}
\hfill
\begin{subfigure}{0.495\textwidth}
    \centering
    \includegraphics[width=\linewidth]{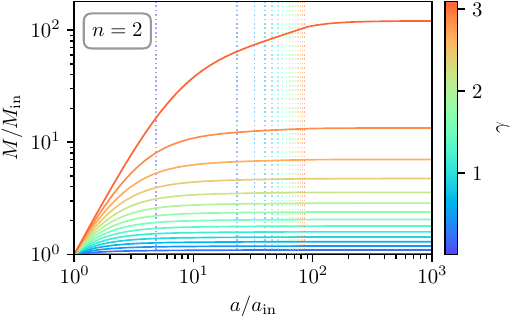}
    \caption{}
\end{subfigure}


\begin{subfigure}{0.51\textwidth}
    \centering
    \includegraphics[width=\linewidth]{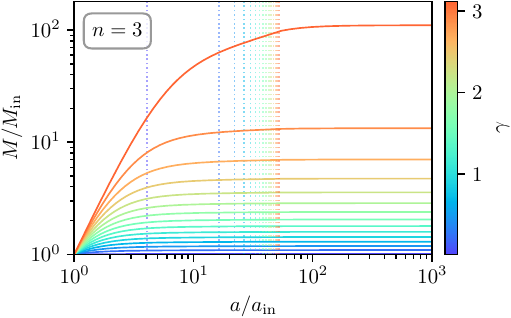}
    \caption{}
\end{subfigure}

\caption{Evolution of the PBH mass ratio $M/M_{\rm in}$ due to accretion as a function of the scale factor $a/a_{\rm in}$. The panels correspond to the background inflaton potentials (a) $n=1$, (b) $n=2$, and (c) $n=3$. The plots are generated using the $\alpha$-attractor parameter $\alpha = 1$ with a fixed reheating temperature $T_{\rm re} = 2 \times 10^{10} \text{ GeV}$ and an initial PBH mass of $ M_{\rm in} = 10^8\,\mathrm{g} $. Because of the expansion history degeneracy for the $n=2$ case, a benchmark value of $n_s = 0.965$ is explicitly chosen to fix the inflationary boundary conditions. The colored solid lines represent the mass evolution for various values of the initial mass fraction $\gamma$, bounded from above by the critical runaway accretion limit $\gamma_c$. The color-matched vertical dotted lines indicate the scale factor at the end of reheating ($a_{\rm re}/a_{\rm in}$) corresponding to each initial mass fraction.}
\label{fig:mass_evolution_cmb}
\end{figure}

To explicitly demonstrate the impact of the radiation bath on the accretion dynamics, we evaluate the PBH mass evolution for initial mass fractions $\gamma$ chosen strictly near their respective critical runaway limits $\gamma_c$. For instance, in a Universe driven by a quartic potential ($n=2$) with an initial mass fraction near the critical limit ($\gamma \simeq 3.10$), pure inflaton accretion limits the total mass growth to $M_{\rm acc}/M_{\rm in} \approx 50$. However, introducing a high reheating temperature of $T_{\rm re} \sim 10^{10.4} \text{ GeV}$ allows the relativistic radiation bath to dominate the late-stage accretion, pushing the final asymptotic mass to $M_{\rm acc}/M_{\rm in} \gtrsim 10^3$---an amplification of the PBH mass by more than a factor of twenty. In Fig.~\ref{fig:radiation_accretion_impact}, we plot the mass ratio $M/M_{\rm in}$ against the scale factor expansion $a/a_{\rm in}$ for an initial mass of $10^8 \, \rm g$, isolating the contribution of radiation accretion by varying the reheating temperature $T_{\rm re}$ across the regime where its effect becomes non-trivial. The black dashed curves represent the theoretical baseline where the PBH accretes exclusively from the oscillating inflaton field. The solid colored curves incorporate the full combined accretion of both the inflaton and the newly produced radiation fluid. As anticipated by the threshold condition in Eq.~\eqref{eq:Tre_min_inequality}, for sufficiently high reheating temperatures (where the radiation energy density builds up more rapidly), the radiation accretion term provides a substantial late-time enhancement to the PBH mass. This enhancement diverges from the pure-inflaton baseline and becomes particularly pronounced just before the end of the reheating epoch (indicated by the vertical dashed lines). Once the universe fully transitions to radiation domination ($a > a_{\rm re}$), the inflaton field is depleted, and the remaining radiation accretion is rapidly quenched by the $a^{-4}$ dilution of the thermal bath. Consequently, the mass quickly converges to its final, enhanced asymptotic plateau.

Having established the conditions under which the radiation bath significantly alters PBH growth, we now examine the broader accretion dynamics across varying initial mass fractions $\gamma$. To effectively highlight this interplay, we fix the initial mass at $10^8 \, \rm g$ and adopt a benchmark reheating temperature of $T_{\rm re} = 2 \times 10^{10} \, \rm GeV$, a scale explicitly chosen because, as demonstrated above, it strongly enhances the combined effects of both inflaton and radiation accretion. Under this specific thermal history, we plot the integrated mass evolution $M/M_{\rm in}$ as a function of the scale factor $a/a_{\rm in}$ in Fig.~\ref{fig:mass_evolution_cmb}. Consistent with our analytical limits, the PBH undergoes a phase of rapid mass growth immediately following its formation. As the universe expands, the background energy densities of both the inflaton and the newly produced radiation bath dilute, causing the accretion rate to become highly suppressed. Ultimately, the mass growth plateaus, capping the PBH at its asymptotic maximum mass $M_{\rm acc}$. To map this accretion history onto the cosmological timeline, we have included color-matched vertical dotted lines indicating the end of the reheating epoch ($a_{\rm re}/a_{\rm in}$) for each respective initial mass fraction $\gamma$. As clearly demonstrated, for the vast majority of the parameter space, the PBH successfully reaches its asymptotic maximum mass well before the universe fully transitions to radiation domination.

It is important to explicitly state that during the reheating epoch, the background energy densities are extreme. In a fully coupled Einstein-Klein-Gordon system, the infalling scalar field would inevitably induce metric backreaction, leading to a dynamic horizon rather than a strictly fixed background. The accretion dynamics derived in this work rely on a quasi-static test-field approximation. By solving the Klein-Gordon equation on a fixed Schwarzschild background and upgrading the mass to a time-dependent variable $M(t)$ a posteriori, our framework effectively serves as an analytical approximation. This approach relies on the rapid separation of scales shortly after PBH formation from gravitational collapse, where $r_s \ll H^{-1}$. In this regime, we can define a dimensionless parameter $\eta \equiv r_s H \ll 1$. Furthermore, we can define a characteristic length scale, $\tilde{r}$, which marks the boundary of the BH's sphere of influence. This scale is found by comparing the BH mass $M$ with the enclosed background energy, yielding $M \simeq \frac{4}{3}\pi\tilde{r}^3\rho$. Physically, this means that up to the distance $\tilde{r}$, the BH mass-energy equivalent strictly dominates over the background energy. This relation evaluates to $\tilde{r} \simeq r_s / \eta^{2/3}$. Because $\eta \ll 1$, this sphere of influence extends far beyond the event horizon ($\tilde{r} \gg r_s$). Within this intermediate region, the spacetime is well-approximated by a quasi-static Schwarzschild geometry, where the Hubble flow acts merely as a minor perturbation. Recent advances in 3+1 Numerical Relativity, utilizing frameworks such as the Einstein Toolkit and Cactus, have demonstrated that metric backreaction can meaningfully alter horizon dynamics and wave scattering processes compared to test-field limits~\cite{Loffler:2011ay, Okawa:2014nda, Cardoso:2015fqa}. Numerical investigation we leave for our future studies.

\subsection{Breakdown of the Effective Fluid Approximation}

A common approach in the literature is to bypass the microphysics of the scalar field and instead model the oscillating inflaton as a macroscopic perfect fluid, characterized entirely by its cycle-averaged effective equation of state, $\omega_\Phi = (n-1)/(n+1)$. Under this perfect fluid approximation, the relativistic accretion rate onto a Schwarzschild black hole is given by the standard hydrodynamic relation~\cite{Das:2025vts, Kalita:2025fcs}
\begin{equation}
\frac{dM}{dt} = \frac{\lambda_c}{16 \pi M_p^4} M^2 \rho_{\infty},
\end{equation}
where $\rho_\infty$ is the background energy density of the fluid, and $\lambda_c$ is the dimensionless accretion efficiency parameter, which depends strictly on the equation of state
\begin{equation}
\lambda_c = \frac{1}{4} \omega_\Phi^{-3/2} (1+3\omega_\Phi)^{\frac{1+3\omega_\Phi}{2\omega_\Phi}}.
\end{equation}
Integrating this fluid accretion rate over the cosmic expansion history from the time of formation ($t_{\rm in}$) yields the analytical mass evolution
\begin{equation}
\frac{M(t)}{M_{\rm in}} = \left[ 1 - \frac{\lambda_c \gamma}{2(1+\omega_\Phi)} \left( \frac{\frac{3}{2}H_{\rm in}(1+\omega_\Phi)(t-t_{\rm in})}{\frac{3}{2}H_{\rm in}(1+\omega_\Phi)(t-t_{\rm in})+1} \right) \right]^{-1}.
\label{eq:acc_mass_evolution_fluid}
\end{equation}
However, replacing the wave dynamics of the inflaton with an effective macroscopic fluid vastly overestimates the efficiency of the mass accretion. To illustrate this discrepancy, consider a PBH forming with an initial mass fraction of $\gamma = 0.2$ in a universe dominated by a quartic potential ($n=2$). Macroscopically, this field behaves as radiation ($\omega_\Phi = 1/3$). For a pure radiation fluid, the accretion efficiency is highly peaked ($\lambda_c \simeq 10.4$), and the fluid approximation predicts a massive asymptotic mass growth of $M_{\rm acc}/M_{\rm in} \simeq 4.53$. In stark contrast, our exact field-level derivation using the matched Klein-Gordon solutions for $n=2$ and $\gamma = 0.2$ yields a much more suppressed asymptotic growth of $M_{\rm acc}/M_{\rm in} \simeq 1.4$. 

\begin{figure}[t]
    \centering
    \includegraphics[width=0.55\linewidth]{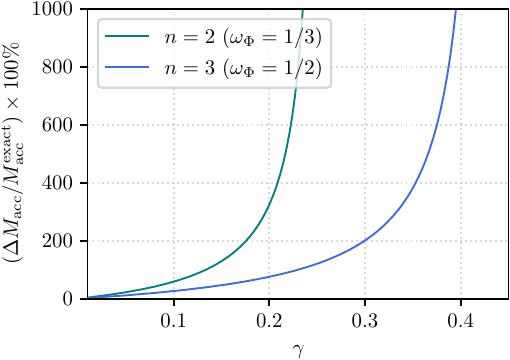}
    \caption{Percentage overestimation of the final accreted PBH mass when using the macroscopic effective fluid approximation compared to our exact scalar field calculation. The solid lines represent the relative error for background inflaton potentials with $n=2$ ($\omega_\Phi = 1/3$, teal) and $n=3$ ($\omega_\Phi = 1/2$, blue). The initial mass fraction $\gamma$ is varied up to the critical runaway limit of the respective effective fluid model. The severe divergence at larger values of $\gamma$ highlights the failure of the fluid approximation to account for horizon-scale wave scattering, confirming the necessity of a rigorous field-level treatment.}
    \label{fig:accretion_error}
\end{figure}

This severe overestimation occurs because the perfect fluid approximation fundamentally breaks down at the horizon scale for an oscillating scalar field~\cite{Wang:2026sgz}. The hydrodynamic equation assumes a continuous, pressure-driven flow of particles falling directly into the singularity. In reality, the scalar field possesses an inherent wave nature. The accretion process is highly subject to scattering off the gravitational potential barrier, and the field cannot efficiently funnel into the black hole if its Compton wavelength or oscillation scale is comparable to the Schwarzschild radius.

To quantitatively illustrate this severe discrepancy, we compare the asymptotic accreted mass derived from our exact field-level calculation, $M_{\rm acc}^{\rm exact}$, (Eq.~\eqref{eq:M_acc}) against the prediction from the effective fluid model, $M_{\rm acc}^{\rm eff} = [1-\lambda_c \gamma / 2(1+\omega_\Phi)]^{-1}$. We define the percentage overestimation as $(M_{\rm acc}^{\rm eff} - M_{\rm acc}^{\rm exact}) / M_{\rm acc}^{\rm exact} \times 100\%$. In Fig.~\ref{fig:accretion_error}, we plot this percentage error as a function of the initial mass fraction $\gamma$ for universes dominated by quartic ($n=2$, $\omega_\Phi=1/3$) and sextic ($n=3$, $\omega_\Phi=1/2$) inflaton potentials. The parameter $\gamma$ is varied up to the critical runaway limit predicted by the effective fluid equation. For very small initial mass fractions ($\gamma \ll 0.1$), where accretion is generally inefficient, the fluid approximation introduces a modest, yet non-negligible, error. However, as $\gamma$ increases toward the critical runaway limit, the effective fluid model's failure to account for wave scattering of the black hole effective potential barrier. This leads to a catastrophic overestimation of the accreted mass, quickly exceeding hundreds of percent as $\gamma$ approaches the threshold. 

This divergence underscores a fundamental conclusion: while the effective equation of state $\omega_\Phi$ successfully captures the cosmological background expansion, it completely fails to capture the local, microphysical wave dynamics required for accurate accretion modeling. Therefore, a rigorous field-level matching methodology, as derived in this work, is absolutely essential to prevent unphysical divergences in PBH mass tracking.


\section{Hawking Evaporation and PBH Lifetime}
\label{sec:hawking_evaporation}

While the early life cycle of a PBH is dominated by rapid mass accretion from the dense surrounding environment, its late-time evolution is governed by mass loss via Hawking radiation. 


\paragraph{Particle Emission and Mass Decay Rate:} The particle emission rate per unit volume from an uncharged, non-rotating (Schwarzschild) PBH for a given particle species $i$ is described by the Bose-Einstein or Fermi-Dirac distribution~\cite{Hawking:1975vcx}
\begin{equation}
n_{E_i} = \frac{\Gamma_{s_i}(E_i)}{e^{8\pi G M E_i} - (-1)^{2s_i}},
\end{equation}
where $M$ is the PBH mass, $s_i$ is the spin of the species, and $E_i = \sqrt{\mu_i^2 + p^2}$ is the energy of the emitted particle with mass $\mu_i$ and momentum $p$. The particle emission rate per unit energy interval is given by
\begin{equation}
\frac{d^2 N_i}{dE_i dt} = \frac{g_i}{2\pi} \frac{\Gamma_{s_i}(E_i)}{e^{8\pi G M E_i} - (-1)^{2s_i}},
\end{equation}
where $g_i$ represents the internal degrees of freedom of the $i^{\text{th}}$ species. Changing variables from energy to momentum ($E_i dE_i = p dp$) and expressing the greybody factor $\Gamma_{s_i}(E_i)$ in terms of the absorption cross-section $\sigma_{s_i}(E_i) = \Gamma_{s_i}(E_i) \pi / p^2$, we obtain
\begin{equation}
\frac{d^2 N_i}{dp dt} = \frac{g_i}{2\pi^2} \frac{\sigma_{s_i}(E_i)}{e^{8\pi G M E_i} - (-1)^{2s_i}} \frac{p^3}{E_i}.
\end{equation}
In the high-energy geometrical optics limit, the cross-section approaches $\sigma_{s_i}(E_i) \simeq 27\pi G^2 M^2$. Defining a normalized greybody factor $\varphi_{s_i}(E_i) = \sigma_{s_i}(E_i) / (27\pi G^2 M^2)$, the momentum spectrum becomes
\begin{equation}
\frac{d^2 N_i}{dp dt} = \frac{g_i}{2\pi^2} \frac{27\pi G^2 M^2 \varphi_{s_i}(E_i)}{e^{8\pi G M E_i} - (-1)^{2s_i}} \frac{p^3}{E_i}.
\label{eq:momentum_spectrum}
\end{equation}
The total mass decay rate is found by integrating the energy-weighted spectrum over all emitted species~\cite{Page:1976df}
\begin{equation}
\frac{dM}{dt} = -\sum_{i} \int_0^{\infty} E_i \frac{d^2 N_i}{dp dt} dp = -\frac{27\pi G^2 M^2}{2\pi^2} \sum_{i} g_i \int_{\mu_i}^{\infty} \frac{\varphi_{s_i}(E_i) (E_i^2 - \mu_i^2)}{e^{8\pi G M E_i} - (-1)^{2s_i}} E_i dE_i.
\end{equation}
Introducing the dimensionless variables $x = 8\pi G M E_i = E_i / T_{\rm BH}$ and $z_i = \mu_i / T_{\rm BH}$, where $T_{\rm BH} = 1 / (8\pi G M)$ is the Hawking temperature, the integral simplifies to~\cite{Cheek:2021odj, Kalita:2025foa, Chatterjee:2025wnt, Auffinger:2022khh, Arbey:2019jmj, Ewasiuk:2025dwn, Ukwatta:2015iba, MacGibbon:1990zk}
\begin{equation}
\frac{dM}{dt} = -\frac{27 M_p^4}{128 \pi^3 M^2} \sum_{i} g_i \int_{z_i}^{\infty} \frac{\varphi_{s_i}(x) (x^2 - z_i^2)}{e^{x} - (-1)^{2s_i}} x dx \equiv -\epsilon \frac{M_p^4}{M^2},
\end{equation}
where $\epsilon = \sum_i g_i \epsilon_i$ parameterizes the total integrated emission efficiency. Assuming the geometrical optics limit ($\varphi_{s_i}(x) \to 1$) and predominantly massless emission ($z_i \to 0$), $\epsilon$ can be analytically evaluated as $\epsilon \simeq \frac{27}{4} \frac{\pi}{480} g_{\ast}(T_{\rm BH})$, where $g_{\ast}(T_{\rm BH})$ is the effective number of relativistic degrees of freedom at the BH temperature. Integrating the mass decay equation $\int M^2 dM = -\epsilon M_p^4 \int dt$ from the onset of evaporation ($t_{\rm in}$) yields the evaporation mass history
\begin{equation}
M(t) \simeq M_{\rm in} \left[ 1 - \frac{3 \epsilon M_p^4}{M_{\rm in}^3} (t - t_{\rm in}) \right]^{1/3}.
\end{equation}
The total evaporation timescale is effectively $t_{\rm ev} \simeq M_{\rm in}^3 / (3 \epsilon M_p^4)$.


\paragraph{Total Mass Evolution with Accretion and Evaporation:} 
The complete evolution of the PBH mass is governed by the sum of the accretion and evaporation rates. However it is important to note that simply adding the classical and quantum stress-energy tensors is an idealized semi-classical limit. In a realistic, high-density reheating environment, an incoming classical scalar field could interact with the horizon's quantum state, potentially leading to non-linear cross-couplings. 
However, the use of this linear balance equation is physically justified in our model due to an extreme separation of timescales. As derived in the preceding sections, mass accretion is highly efficient immediately after PBH formation but is rapidly quenched by cosmological expansion. During this early accretion-dominated epoch, the PBH evaporate is suppressed rendering 
the associated quantum evaporation rate entirely negligible. Conversely, Hawking evaporation only becomes dynamically explosive at the end of the PBH's life, long after the classical inflaton field has decayed and the surrounding environment has diluted into a sparse thermal bath. Because the classical accretion flux and the quantum evaporation flux do not significantly overlap in time, any non-linear cross-coupling is dynamically suppressed, allowing us to safely decouple the differential equations. The PBH quickly accretes to its asymptotic maximum mass $M_{\text{acc}}$ (Eq.~\ref{eq:M_acc}), which subsequently serves as the effective initial mass for the strictly late-time evaporation phase.

To track the evaporation against the cosmological background, we must map the time $t$ to the scale factor $a$. The background scale factor evolves differently depending on whether the PBH completely evaporates during the reheating era ($t_{\rm ev} < t_{\rm re}$) or survives into the radiation-dominated era ($t_{\rm ev} > t_{\rm re}$). Using the scaling relations $t \propto a^{\frac{3}{2}(1+\omega_\Phi)}$ during reheating and $t \propto a^2$ during radiation domination ($\omega_r = 1/3$), we map the time ratio $t/t_{\rm in}$:
\begin{equation}
\frac{t}{t_{\rm in}} = 
\begin{cases}
\left( \frac{a}{a_{\rm in}} \right)^{\frac{3}{2}(1+\omega_\Phi)} & \text{for } t_{\rm ev} < t_{\rm re}, \\
\left( \frac{a}{a_{\rm in}} \right)^{2} \left\{ \frac{(5-3\omega_\Phi)}{2(1+\omega_\Phi)} \frac{4 \pi \gamma M_p^2}{\Gamma_\Phi M_{\rm in}} \right\}^{\frac{\left(3 \omega_\Phi -1\right) }{3(1+\omega_\Phi)}} & \text{for } t_{\rm ev} > t_{\rm re}.
\end{cases}
\end{equation}
Substituting this time-to-scale-factor mapping into the evaporation equation, and replacing the formation mass with the fully accreted mass ($M_{\rm in} \to M_{\rm acc}$), the final phenomenological mass evolution of the PBH is 
\begin{equation}
M(a) \approx M_{\rm acc}
\begin{cases}
\left[ 1 - \frac{3 \epsilon M_p^4 t_{\rm in}}{M_{\rm acc}^3} \left( \left( \frac{a}{a_{\rm in}} \right)^{\frac{3}{2}(1+\omega_\Phi)} - 1 \right) \right]^{1/3} & \text{if } t_{\rm ev} < t_{\rm re}, \\
\left[ 1 - \frac{3 \epsilon M_p^4 t_{\rm in}}{M_{\rm acc}^3 } \left( \left( \frac{a}{a_{\rm in}} \right)^{2} \left\{ \frac{(5-3\omega_\Phi)}{2(1+\omega_\Phi)} \frac{4 \pi \gamma M_p^2}{\Gamma_\Phi M_{\rm acc}} \right\}^{\frac{\left(3 \omega_\Phi - 1\right)}{3(1+\omega_\Phi)}} - 1 \right) \right]^{1/3} & \text{if } t_{\rm ev} > t_{\rm re}.
\end{cases}
\end{equation}
\begin{figure}[t]
\centering

\begin{subfigure}{0.51\textwidth}
    \centering
    \includegraphics[width=\linewidth]{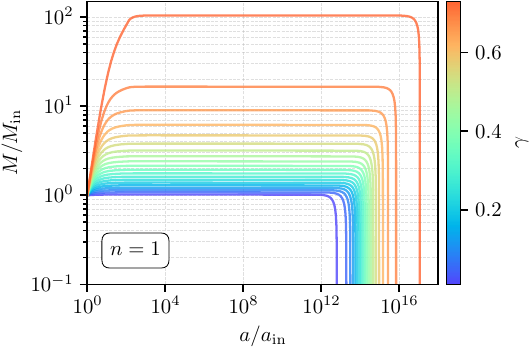}
    \caption{}
\end{subfigure}
\hfill
\begin{subfigure}{0.48\textwidth}
    \centering
    \includegraphics[width=\linewidth]{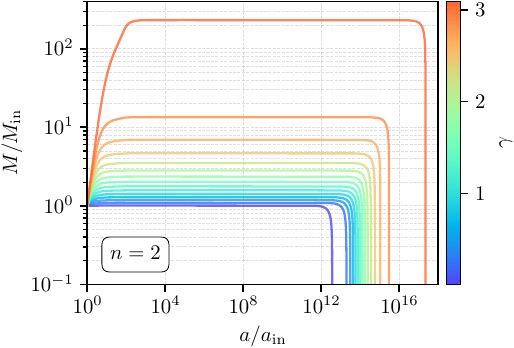}
    \caption{}
\end{subfigure}
\begin{subfigure}{0.51\textwidth}
    \centering
    \includegraphics[width=\linewidth]{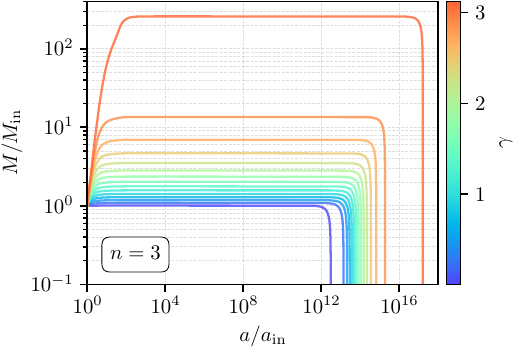}
    \caption{}
\end{subfigure}
\caption{Evolution of the PBH mass ratio $M/M_{\rm in}$, accounting for both accretion and evaporation, as a function of the scale factor $a/a_{\rm in}$. The panels correspond to different background inflaton potentials: (a) $ n = 1 $, (b) $ n = 2 $, and (c) $ n = 3 $. The results are shown for the $\alpha$-attractor parameter $ \alpha = 1 $, with a fixed reheating temperature $ T_{\rm re} = 2\times 10^{10}\,\mathrm{GeV} $ and an initial PBH mass $ M_{\rm in} = 10^8\,\mathrm{g} $. The colored solid curves depict the evolution for different values of the initial mass fraction $ \gamma $, up to the critical threshold $ \gamma_c $ beyond which runaway accretion occurs.}
\label{fig:total_mass_evolution}
\end{figure}
Whether a PBH evaporates entirely within the reheating epoch or survives into the radiation-dominated universe depends on its initial mass. The boundary condition $t_{\rm ev} > t_{\rm re}$ requires 
\begin{equation}
\frac{M_{\rm in}^{3}}{3 \epsilon M_p^4} > t_{\rm in} \left( \frac{a_{\rm re}}{a_{\rm in} } \right)^{\frac{3}{2}(1+\omega_\Phi)}.
\end{equation}
Substituting the initial formation time $t_{\rm in} = M_{\rm in} / [6\pi \gamma M_p^2 (1+\omega_\Phi)]$ and the geometric expansion ratio $(a_{\rm re}/a_{\rm in})^{\frac{3}{2}(1+\omega_\Phi)}$, the inequality simplifies algebraically 
\begin{equation}
\frac{M_{\rm in}^{3}}{3 \epsilon M_p^4} > \left[ \frac{M_{\rm in}}{6\pi \gamma M_p^2 (1+\omega_\Phi)} \right] \left[ \frac{5-3\omega_\Phi}{2(1+\omega_\Phi)} \frac{4\pi \gamma M_p^2}{\Gamma_\Phi M_{\rm in}} \right] = \frac{5-3\omega_\Phi}{3(1+\omega_\Phi)^2 \Gamma_\Phi}.
\end{equation}
Isolating $M_{\rm in}$, we find the critical initial mass required for a PBH to survive the reheating epoch 
\begin{equation}
M_{\rm in} > M_p \left\{ \epsilon \frac{(5-3\omega_\Phi)}{(1+\omega_\Phi)^2} \frac{M_p}{\Gamma_\Phi} \right\}^{1/3}.
\label{eq:evaporation_survival}
\end{equation}
PBHs forming with masses below this threshold will evaporate entirely while the universe is still dominated by the oscillating inflaton field. Assuming standard Standard Model degrees of freedom during evaporation ($g_*(T_{\rm BH}) \approx 108 \implies \epsilon \approx 4.77$) and at reheating ($g_{*,{\rm re}} \approx 106.75$), we can evaluate this critical boundary. For high-scale reheating ($T_{\rm re} \sim 10^{15} \text{ GeV}$), the epoch is extremely brief, allowing even sub-gram PBHs ($M_{\rm in} \gtrsim 10^{-3} \text{ g}$) to survive into radiation domination. Conversely, for prolonged reheating scenarios ($T_{\rm re} \sim 10^5 \text{ GeV}$), only PBHs heavier than a few kilograms ($M_{\rm in} \gtrsim 5 \times 10^3 \text{ g}$) can withstand the Hawking evaporation long enough to reach the onset of the thermal universe.

\begin{figure}[t]
\centering

\begin{subfigure}{0.49\textwidth}
    \centering
    \includegraphics[width=\linewidth]{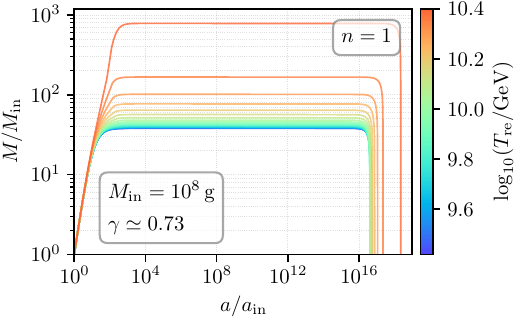}
    \caption{}
\end{subfigure}
\hfill
\begin{subfigure}{0.49\textwidth}
    \centering
    \includegraphics[width=\linewidth]{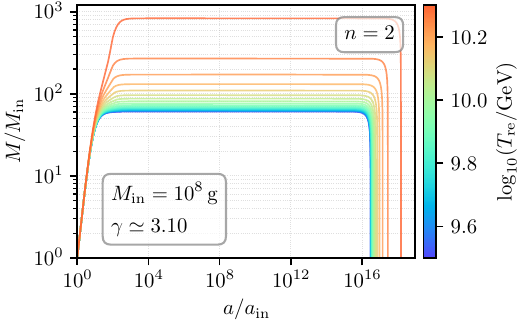}
    \caption{}
\end{subfigure}


\begin{subfigure}{0.51\textwidth}
    \centering
    \includegraphics[width=\linewidth]{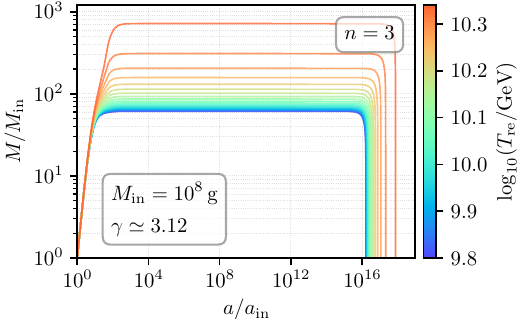}
    \caption{}
\end{subfigure}

\caption{Complete lifecycle evolution of the PBH mass ratio $M/M_{\rm in}$ as a function of the scale factor $a/a_{\rm in}$, highlighting the effect of the reheating temperature $T_{\rm re}$. The panels represent background inflaton potentials with (a) $n=1$, (b) $n=2$, and (c) $n=3$. The initial PBH mass is fixed at $M_{\rm in} = 10^8 \text{ g}$, with initial mass fractions $\gamma$ chosen near their critical limits ($\gamma \simeq 0.73, 3.10,$ and $3.12$, respectively). The color map denotes the varying reheating temperature $\log_{10}(T_{\rm re}/\text{GeV})$. Higher values of $T_{\rm re}$ enhance the late-time radiation accretion, slightly elevating the plateau mass. Due to the cubic dependence of the evaporation timescale on the PBH mass, this enhancement significantly prolongs the PBH lifetime, shifting the final evaporation epoch to exponentially larger scale factors.}
\label{fig:total_mass_evolution_Tre}
\end{figure}

By synthesizing our analytical models for both the early-time accretion and the late-time Hawking radiation, we can now track the complete mass evolution of a PBH from its formation to its ultimate evaporation. In Fig.~\ref{fig:total_mass_evolution}, we plot the full mass ratio $M/M_{\rm in}$ against the scale factor expansion $a/a_{\rm in}$. The dynamics clearly separate into three distinct phenomenological phases. First, the PBH undergoes a brief epoch of rapid mass growth driven by the dense scalar and radiation fields, reaching its asymptotic maximum mass $M_{\rm acc}$. Second, an extended plateau phase emerges, during which the cosmological expansion has sufficiently diluted the background fields to halt accretion, while the PBH is still too massive for Hawking evaporation to be efficient. Finally, as the slow mass loss gradually reduces the PBH's size, the evaporation rate $dM/dt \propto -1/M^2$ triggers a runaway process, leading to a sudden, catastrophic loss of mass characterized by the steep vertical drop-off in the plot.

Furthermore, the precise thermal history of the universe plays a critical role in determining the ultimate lifespan of these BHs. To illustrate this, we present Fig.~\ref{fig:total_mass_evolution_Tre}, which tracks the complete lifecycle for a fixed initial mass of $M_{\rm in} = 10^8 \, \text{g}$ and initial mass fractions $\gamma$ set just below their respective critical runaway limits. By varying the reheating temperature $T_{\rm re}$ (indicated by the color map), we observe a compounding effect on the PBH lifetime. As established in Section \ref{sec:radiation_accretion}, higher reheating temperatures lead to a more substantial contribution from radiation accretion prior to the end of the reheating epoch, thereby slightly increasing the asymptotic plateau mass $M_{\rm acc}$. Because the Hawking evaporation timescale is proportional to the cube of this maximum mass ($t_{\rm ev} \propto M_{\rm acc}^3$), even modest enhancements in accretion due to a hotter thermal bath result in a drastic delay of the final evaporation epoch. Consequently, the vertical drop-off marking the end of the PBH's life shifts to significantly larger scale factors, extending its survival much deeper into the radiation-dominated era.

\begin{figure}[t]
\centering
\includegraphics[width=0.6\linewidth]{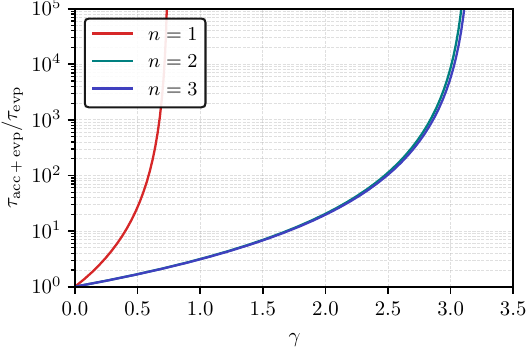} 
\caption{Enhancement of the total PBH lifetime due to the initial accretion phase as a function of the initial mass fraction $\gamma$. The plot shows the ratio of the combined accretion-evaporation lifetime ($\tau_{\text{acc+evp}}$) to the lifetime of a PBH with the same initial mass that only experiences evaporation ($\tau_{\text{evp}}$). The curves correspond to inflaton potentials with $n=1$ (red), $n=2$ (teal), and $n=3$ (blue). Due to the $t_{\text{ev}} \propto M^3$ scaling of the evaporation timescale, the total lifetime increases by many orders of magnitude as $\gamma$ approaches the critical runaway accretion limit $\gamma_c$.}
\label{fig:lifetime_enhancement}
\end{figure}

To more directly quantify the massive increase in PBH lifespan described by the cubic mass dependence of the evaporation timescale, we present Fig.~\ref{fig:lifetime_enhancement}. This plot shows the ratio of the total PBH lifetime, accounting for both accretion and evaporation ($\tau_{\text{acc+evp}}$), to the lifetime the same BH would have if it only evaporated ($\tau_{\text{evp}}$). In our separate-timescale approximation, this ratio can be simplified to leading order as 
\begin{equation}
    \frac{\tau_{\text{acc+evp}}}{\tau_{\text{evp}}} \approx \left( \frac{M_{\text{acc}}}{M_{\text{in}}} \right)^3 \simeq \left\{ 1 - \frac{2\pi}{3} \gamma \, n(n+1) \left( \mathcal{P}_1^{(n)} \frac{\Gamma\left(\frac{1}{2n} + \frac{1}{2}\right)}{\Gamma\left(\frac{1}{2n}\right)} \right)^2 \right\}^{-3} .
\end{equation}
As expected, for small values of $\gamma$, where accretion is negligible ($M_{\text{acc}} \approx M_{\text{in}}$), the ratio is close to unity. However, as $\gamma$ increases and approaches the critical runaway accretion limit $\gamma_c$, the ratio explodes by multiple orders of magnitude. For example, for $\gamma$ just below the threshold, the PBH can live $10^5$ times longer than it would without accretion. This extreme sensitivity is a direct result of the rapid initial accretion pushing the BH mass towards a singularity, which then dramatically delays the start of the efficient evaporation phase.


\section{Gravitational Wave Production from Evaporation}
\label{sec:gravitational_waves}

As the PBHs evaporate, they emit all particle species present in the standard model, as well as any beyond-the-standard-model particles, including gravitons. The emission of these gravitons generates a stochastic gravitational wave background (SGWB) that persists to the present day. The graviton production rate per unit time and per unit angular frequency $\omega$ from a single Schwarzschild BH is given by the instantaneous Hawking spectrum~\cite{PhysRevD.14.3260}
\begin{equation}
Q_{\rm GW}(t, \omega) = \frac{d^2 N_{\rm grav}}{dt d\omega} = \sum_{\ell,m} \frac{\Gamma_{\ell m}^{(s=2)}(\omega)}{e^{\omega /T_{\rm BH}} - 1}.
\end{equation}
It is important to emphasize that the production of these gravitational waves does not violate Birkhoff's theorem. Classically, a strictly spherically symmetric system cannot emit gravitational waves, as such emission requires a time-varying quadrupole moment. However, the Stochastic Gravitational Wave Background (SGWB) computed here originates entirely from quantum Hawking evaporation. Quantum fluctuations of the tensor field exist in the vacuum surrounding the black hole and are converted into real gravitons near the event horizon. Because this quantum particle production occurs within a spherically symmetric classical background, the resulting emission of gravitons is isotropic. Consequently, the emitted radiation forms a stochastic background that preserves the macroscopic symmetries of the spacetime, conceptually analogous to the generation of a stochastic gravitational wave background from quantum vacuum fluctuations during cosmological inflation.

To carefully account for the graviton flux, including the exact greybody factors $\Gamma_{\ell m}^{(s)}$, we compute $Q_{\rm GW}(t, \omega)$ numerically using the public code \texttt{BlackHawk}\footnote{Available at \href{https://blackhawk.hepforge.org}{https://blackhawk.hepforge.org}}~\cite{Arbey:2019mbc, Arbey:2021mbl}. The total differential energy density of the emitted gravitational waves is therefore weighted by the PBH number density $n_{\rm BH}(t)$. Evaluated per logarithmic frequency interval, this yields 
\begin{equation}
\frac{d^2\rho_{\rm GW}}{dt d\ln\omega} = n_{\rm BH}(t) \frac{\omega^2}{2\pi} Q_{\rm GW}(t, \omega).
\end{equation}
To relate the GW energy density at the time of emission to the present day, we must track the cosmological evolution of each component. The physical frequency redshifts as $\omega \propto a^{-1}$, the PBH number density dilutes as $n_{\rm BH} \propto a^{-3}$ (assuming a matter-like PBH population prior to evaporation), and the GW energy density dilutes as radiation, $\rho_{\rm GW} \propto a^{-4}$. Defining quantities evaluated at the end of evaporation with the subscript ``ev'' and at the time of formation with ``in'', we can express the differential energy density evaluated at the end of evaporation as 
\begin{equation}
\frac{d\rho_{\rm GW}^{\rm ev}}{dt d\ln\omega_{\rm ev}} = n_{\rm BH}^{\rm in} \frac{\omega_{\rm ev}^2}{2\pi} \left( \frac{a_{\rm in}}{a_{\rm ev}} \right)^3 \left( \frac{a_{\rm ev}}{a} \right) Q_{\rm GW}\left(t, \omega_{\rm ev} \left(\frac{a}{a_{\rm ev}}\right)^{-1}\right).
\end{equation}
Integrating this emission over the entire lifetime of the BH gives the total accumulated GW energy density at $t = t_{\rm ev}$ 
\begin{equation}
\frac{d\rho_{\rm GW}^{\rm ev}}{d\ln\omega_{\rm ev}} = n_{\rm BH}^{\rm in} \frac{\omega_{\rm ev}^2}{2\pi} \left( \frac{a_{\rm in}}{a_{\rm ev}} \right)^3 \int_{t_{\rm in}}^{t_{\rm ev}} dt \left( \frac{a}{a_{\rm ev}} \right)^{-1} Q_{\rm GW}\left(t, \omega_{\rm ev} \left(\frac{a}{a_{\rm ev}}\right)^{-1}\right).
\end{equation}
To find the observable abundance today, we redshift the energy density from $a_{\rm ev}$ to $a_0 = 1$. The present-day GW energy density is $d\rho_{\rm GW}^0 / d\ln\omega_0 = (d\rho_{\rm GW}^{\rm ev} / d\ln\omega_{\rm ev}) (a_{\rm ev}/a_0)^4$, and the observed frequency is $f_0 = \omega_0 / 2\pi = (\omega_{\rm ev}/2\pi) (a_{\rm ev}/a_0)$. The present GW relic abundance parameter, defined as the energy density fraction relative to the critical density $\rho_{\rm crit,0} = 3 M_p^2 H_0^2$, evaluates to~\cite{Anantua:2008am, Ireland:2023avg}
\begin{equation}
\Omega_{\rm GW} (f_0) = \frac{n_{\rm BH}^{\rm in} \omega_0^2}{6 \pi M_p^2 H_0^2} \frac{a_{\rm in}^3}{a_0^2} \int_{t_{\rm in}}^{t_{\rm ev}} \frac{dt}{a} \, Q_{\rm GW}\left(t, \frac{\omega_0 a_0}{a}\right).
\end{equation}
We parameterize the initial PBH number density in terms of the initial mass fraction $\beta$, representing the ratio of the PBH energy density to the critical density at the time of formation as
\begin{equation}
\beta = \frac{\rho_{\rm BH}^{\rm in}}{\rho_{\rm crit}^{\rm in}} = \frac{n_{\rm BH}^{\rm in} M_{\rm in}}{3 M_p^2 H_{\rm in}^2}.
\end{equation}
Substituting the initial horizon mass relation $M_{\rm in} = 4 \pi \gamma M_p^2 H_{\rm in}^{-1}$, we can invert this to find the initial number density 
\begin{equation}
n_{\rm BH}^{\rm in} = 48 \pi^2 \gamma^2 \frac{M_p^6}{ M_{\rm in}^3} \beta .
\end{equation}
Converting the time integral to an integral over the scale factor ($dt = da / (aH)$) and expressing the abundance scaled by the reduced Hubble parameter $h$, the final expression for the SGWB spectrum is 
\begin{equation}
\Omega_{\rm GW}h^2 (f_0) = \frac{24 \pi \gamma^2 M_p^6 \omega_0^2}{\rho_{\rm crit,0}h^{-2} M_{\rm in}^3} \beta a_{\rm in}^3 \int_{a_{\rm in}}^{a_{\rm ev}} \frac{da}{a^2 H(a)} Q_{\rm GW}\left(M(a), \frac{\omega_0}{a}\right),
\end{equation}
where $\rho_{\rm crit,0}h^{-2} \simeq 8.056 \times 10^{-47}\text{ GeV}^4$. Because the integral spans the reheating and radiation-dominated epochs, the Hubble parameter must be evaluated piecewise: $H(a) \propto a^{-\frac{3}{2}(1+\omega_\Phi)}$ for $a < a_{\rm re}$, and $H(a) \propto a^{-2}$ for $a > a_{\rm re}$. The initial scale factor $a_{\rm in}$ is anchored to the present day by tracking back from matter-radiation equality ($z_{\rm eq} \simeq 3400$, $T_{\rm eq} \simeq 0.8 \text{ eV}$):
\begin{equation}
a_{\rm in} = \frac{1}{1+z_{\rm eq}} \left( \frac{T_{\rm eq}}{T_{\rm re}} \right) \left( \frac{H_{\rm in}}{H_{\rm re}} \right)^{-\frac{2}{3(1+\omega_\Phi)}},
\end{equation}
where the Hubble parameter at the end of reheating is defined by the thermal bath 
\begin{equation}
H_{\rm re} = \left( \frac{2\pi^2}{90} g_* \right)^{\frac{1}{2}} \left( \frac{T_{\rm re}}{M_p} \right)^2 M_p.
\end{equation}
For the numerical integration, we supply the full, combined accretion-evaporation mass history $M(a)$ derived in Section \ref{sec:hawking_evaporation} as the dynamic input for the \verb|BlackHawk_inst| script, ensuring the instantaneous graviton flux accurately reflects the evolving size and temperature of the PBH at every time step.

\begin{figure}[t]
\centering

\begin{subfigure}{0.49\textwidth}
    \centering
    \includegraphics[width=\linewidth]{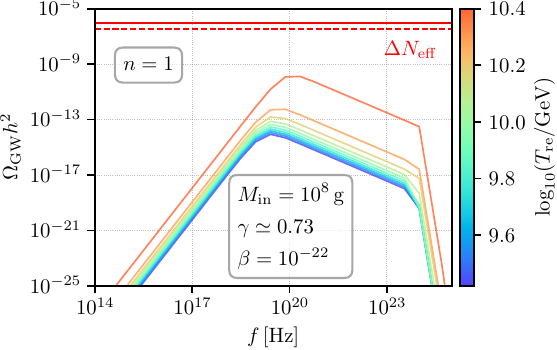}
    \caption{}
\end{subfigure}
\hfill
\begin{subfigure}{0.49\textwidth}
    \centering
    \includegraphics[width=\linewidth]{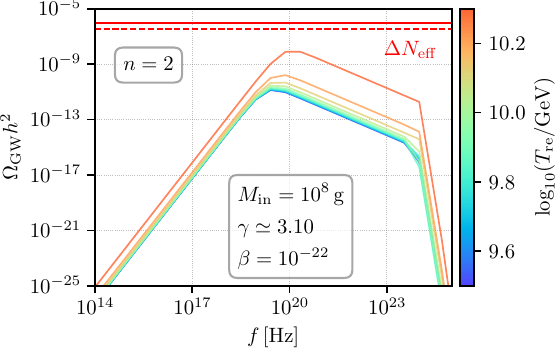}
    \caption{}
\end{subfigure}

\begin{subfigure}{0.51\textwidth}
    \centering
    \includegraphics[width=\linewidth]{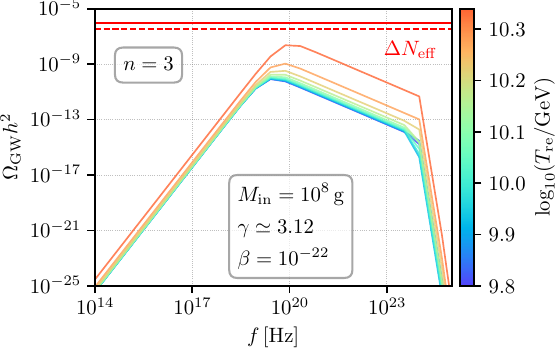}
    \caption{}
\end{subfigure}

\caption{The present-day stochastic gravitational wave background spectrum $\Omega_{\rm GW}h^2$ from PBH evaporation as a function of frequency $f$. The panels correspond to background inflaton potentials with (a) $n=1$, (b) $n=2$, and (c) $n=3$. The initial PBH mass is fixed at $M_{\rm in} = 10^8 \, \text{g}$, and the initial abundance is $\beta = 10^{-22}$. The initial mass fractions $\gamma$ are chosen very close to their critical runaway limits. The colored curves show the enhancement of the GW amplitude as the reheating temperature $T_{\rm re}$ is increased. Higher $T_{\rm re}$ leads to more efficient radiation accretion, extending the PBH lifetime and significantly boosting the GW signal. The horizontal red lines denote the upper bounds from the effective number of relativistic degrees of freedom: the solid line corresponds to $\Delta N_{\rm eff} = 0.17$ (Planck + BAO) \cite{Planck:2018vyg}, and the dashed line corresponds to the projected sensitivity $\Delta N_{\rm eff} = 0.06$ (CMB-S4) \cite{Abazajian:2019eic}.}
\label{fig:gw_spectrum_Tre}
\end{figure}

Having established the complete mass evolution history, we now evaluate the present-day stochastic gravitational wave background (SGWB) produced by the Hawking evaporation of these PBHs. To isolate the observational impact of the radiation accretion phase, we fix the initial PBH mass to $M_{\rm in} = 10^8 \, \text{g}$ and the initial abundance to a conservative benchmark of $\beta = 10^{-22}$. In Fig.~\ref{fig:gw_spectrum_Tre}, we present the SGWB spectra $\Omega_{\rm GW}h^2$ as a function of the present-day frequency $f$. We set the initial mass fraction $\gamma$ just below the critical runaway limit for each respective potential ($n=1, 2, 3$). The color gradient illustrates the effect of varying the reheating temperature $T_{\rm re}$. The spectra exhibit the characteristic shape of Hawking-emitted gravitons: a low-frequency tail scaling as $f^3$, a peak corresponding to the maximum temperature of the BH at the end of its life, and a high-frequency cutoff.

\begin{figure}[t]
\centering
\includegraphics[width=1\textwidth]{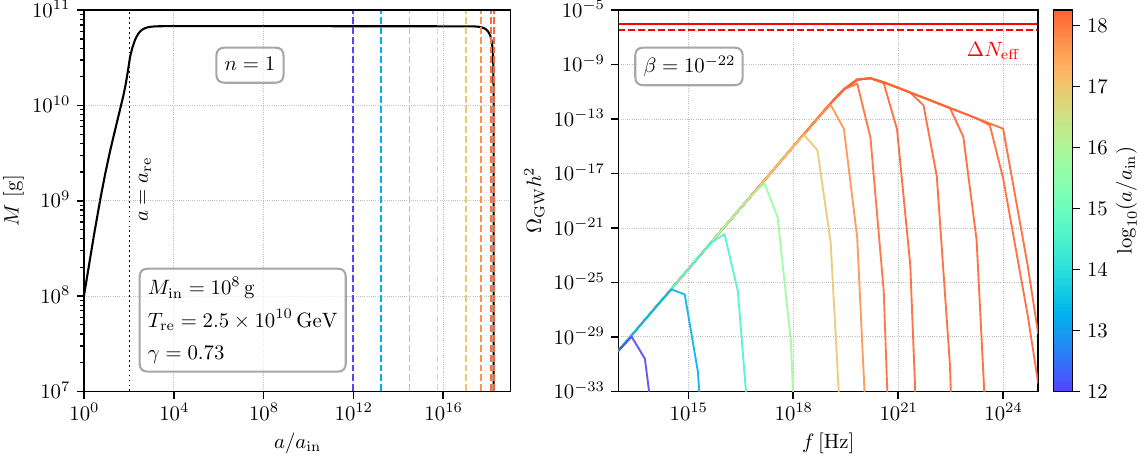}

\caption{Time evolution of the cumulative SGWB spectrum during PBH evaporation. \textbf{Left panel:} Evolution of the PBH mass $M$ as a function of the scale factor $a/a_{\rm in}$ shown for $n=1$ case. The vertical colored dashed lines indicate discrete snapshots during the evaporation phase. \textbf{Right panel:} The cumulative GW spectrum $\Omega_{\rm GW}h^2$ integrated up to the corresponding scale factors marked in the left panel. The color map denotes the scale factor $\log_{10}(a/a_{\rm in})$. Parameters are chosen to maximize the accretion effect: $M_{\rm in} = 10^8 \text{ g}$, $\gamma$ near the critical runaway limit, and $T_{\rm re} = 2.5 \times 10^{10}\,\rm GeV$, with an initial abundance $\beta = 10^{-22}$. Horizontal red lines represent the $\Delta N_{\rm eff}$ bounds from Planck + BAO (solid) and CMB-S4 (dashed).}
\label{fig:gw_time_evolution}
\end{figure}


The most striking feature of these plots is the extreme sensitivity of the peak amplitude to the reheating temperature. As $T_{\rm re}$ increases, the density of the radiation bath during the early accretion phase is higher, which enhances the final accreted mass $M_{\rm acc}$ (as demonstrated in Fig.~\ref{fig:radiation_accretion_impact} and \ref{fig:total_mass_evolution_Tre}). Because the evaporation timescale scales as $t_{\rm ev} \propto M_{\rm acc}^3$, this slight increase in mass dramatically delays the epoch of complete evaporation. Consequently, the PBHs survive longer, their relative energy density compared to the background universe grows larger prior to evaporation, and the emitted gravitational waves suffer less subsequent cosmological dilution. This compounding effect boosts the present-day GW amplitude by multiple orders of magnitude.

Because these ultra-high-frequency gravitational waves act as a form of dark radiation, their integrated energy density is strictly bounded by the effective number of relativistic degrees of freedom, $N_{\rm eff}$, during Big Bang Nucleosynthesis and recombination. We indicate the current upper bound of $\Delta N_{\rm eff} \le 0.17$ from Planck + BAO \cite{Planck:2018vyg} with a solid red line, and the projected sensitivity of $\Delta N_{\rm eff} \le 0.06$ for the future CMB-S4 experiment \cite{Abazajian:2019eic} with a dashed red line. For the chosen parameters, the enhanced GW signal approaches, but safely satisfies, these cosmological bounds, demonstrating that radiation accretion can push otherwise unobservable PBH models into the purview of next-generation constraints.

\begin{figure}[t]
\centering

\begin{subfigure}{0.49\textwidth}
    \centering
    \includegraphics[width=\linewidth]{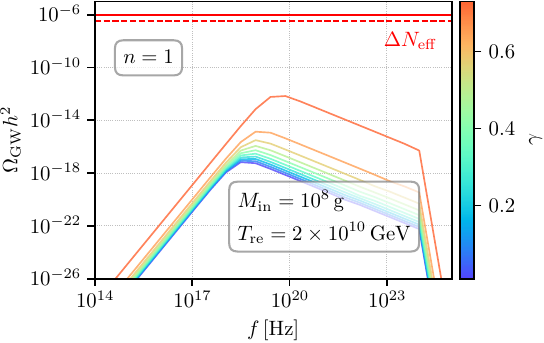}
    \caption{}
\end{subfigure}
\hfill
\begin{subfigure}{0.49\textwidth}
    \centering
    \includegraphics[width=\linewidth]{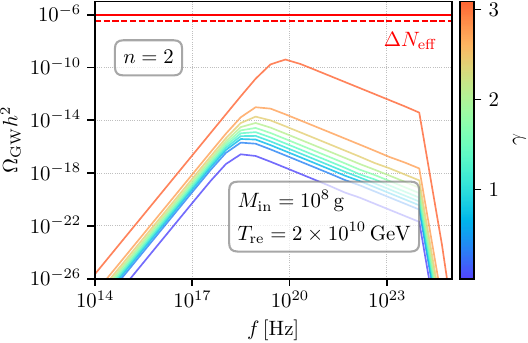}
    \caption{}
\end{subfigure}

\begin{subfigure}{0.51\textwidth}
    \centering
    \includegraphics[width=\linewidth]{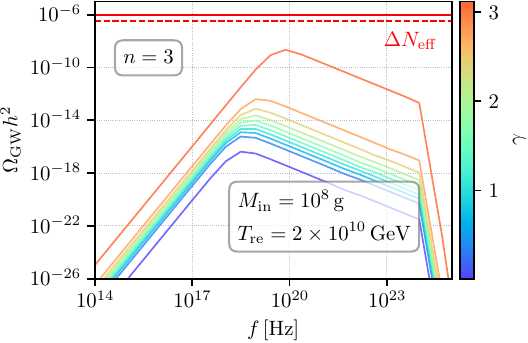}
    \caption{}
\end{subfigure}

\caption{The present-day stochastic gravitational wave background spectrum $\Omega_{\rm GW}h^2$ as a function of frequency $f$, highlighting the effect of the initial PBH mass fraction $\gamma$. The panels correspond to background inflaton potentials with (a) $n=1$, (b) $n=2$, and (c) $n=3$. The initial PBH mass is fixed at $M_{\rm in} = 10^8 \, \text{g}$, and the reheating temperature is set to $T_{\rm re} = 2 \times 10^{10} \, \text{GeV}$ to ensure efficient combined accretion. The color map denotes the value of $\gamma$, which is varied up to the critical runaway accretion limit $\gamma_c$ for each potential. As $\gamma \to \gamma_c$ (orange/red curves), the non-linear enhancement of the fully accreted mass drastically extends the PBH lifetime, resulting in a sudden, multi-order-of-magnitude boost in the gravitational wave amplitude.}
\label{fig:gw_spectrum_gamma}
\end{figure}

To gain deeper physical insight into the generation of this stochastic background, we track the temporal accumulation of the GW spectrum throughout the PBH's lifespan. Because the Hawking temperature is inversely proportional to the BH mass ($T_{\rm BH} \propto 1/M$), the evaporation process is highly non-linear. As the PBH loses mass, its temperature skyrockets, shifting the peak emission to higher frequencies and drastically increasing the total power output. In Fig.~\ref{fig:gw_time_evolution}, we dissect this temporal evolution. The left panel display the PBH mass evolution, with colored vertical dashed lines marking discrete snapshots in time during the evaporation phase for the case of $n=1$. The right panel show the corresponding cumulative SGWB spectrum integrated up to those specific moments. The parameters ($M_{\rm in}$, $T_{\rm re}$, and $\gamma$) are deliberately chosen to optimize the enhanced accretion phase. As clearly demonstrated by the progression from cool (blue) to warm (red) colors, the early stages of the plateau and initial evaporation contribute negligibly to the final amplitude of the spectrum. The high-frequency peak and the vast majority of the overall gravitational wave energy density are generated entirely in the final, explosive moments of the PBH's life (the reddish curves). During this runaway phase, the mass drops precipitously, and the BH rapidly radiates its remaining energy before completely vanishing. Consequently, the observability of the PBH-induced SGWB is overwhelmingly dictated by the extreme physics of its final evaporating moments, making the precise determination of its fully accreted plateau mass $M_{\rm acc}$ absolutely crucial for accurate predictions. Similar behavior is also observed for higher values of $n$.

To complement our analysis of the thermal history, we also examine the sensitivity of the SGWB spectrum to the initial formation parameters of the PBH. In Fig.~\ref{fig:gw_spectrum_gamma}, we evaluate the present-day GW amplitude by fixing the reheating temperature at the optimal value of $T_{\rm re} = 2 \times 10^{10} \, \text{GeV}$ and varying the initial mass fraction $\gamma$ across its allowed parameter space, up to the critical runaway limit $\gamma_c$. The resulting spectra clearly demonstrate the highly non-linear dynamics of the accretion phase. For smaller values of $\gamma$ (represented by the blue and green curves), the accretion is relatively inefficient, leading to moderate enhancements in the final PBH mass and, consequently, a tightly clustered set of lower-amplitude GW spectra. However, as $\gamma$ approaches the critical threshold $\gamma_c$ (the orange and red curves), the initial accretion rate enters the runaway regime. As derived in Section \ref{sec:M_acc}, this runaway accretion pushes the asymptotic maximum mass $M_{\rm acc}$ toward a singularity. Because the evaporation lifetime depends cubically on this mass, the epoch of complete evaporation is delayed by many orders of magnitude in the scale factor. Evaporating much later in the universe's history means the emitted gravitational waves suffer significantly less cosmological redshift and volumetric dilution before reaching the present day. This translates to an explosive, multi-order-of-magnitude boost in the GW energy density, pushing the peak amplitude right up to the $\Delta N_{\rm eff}$ bounds. This confirms that PBHs forming with initial mass fractions near their critical limits will act as overwhelmingly dominant sources of stochastic gravitational waves, strictly constrained by current and future dark radiation limits.


\section{Conclusions}
\label{sec:conclusion}

The life cycle of a PBH is inextricably linked to the cosmological epoch in which it resides. In this paper, we have systematically investigated accretion, and ultimate evaporation of PBHs born during the reheating epoch. By rigorously modeling the background as an oscillating inflaton scalar field with a potential $V \propto \Phi^{2n}$, particularly utilizing the well-motivated $\alpha$-attractor E-model framework, we have tracked the complete, highly non-linear mass evolution of these BHs from their genesis to their explosive demise.

By matching the exact near-horizon solutions of the Klein-Gordon equation to the cosmological far-zone, we demonstrated that the cycle-averaged mass accretion rate of an oscillating scalar field falls off as $t^{-2}$, strictly dictated by the background cosmic expansion. This causes the PBH to undergo an initial phase of rapid, aggressive mass growth that eventually plateaus, capping the BH at an asymptotic maximum mass $M_{\rm acc}$ well before the universe transitions to radiation domination. We showed that treating reheating as a pure scalar-field epoch is insufficient for PBH dynamics. By incorporating the coupled Boltzmann equations for inflaton decay, we found that the accretion of the newly produced, subdominant relativistic thermal bath provides a crucial enhancement to the final PBH mass. This effect is especially pronounced for higher reheating temperatures, where radiation dominates the accretion flow long before it dominates the cosmic energy budget. Because the Hawking evaporation time scales cubically with mass ($t_{\rm ev} \propto M^3$), the initial rapid accretion phase induces a massive change on the PBH's survival. We demonstrated that for initial mass fractions $\gamma$ approaching the critical runaway limit $\gamma_c$, the PBH life span is extended by many orders of magnitude. 

Tracking the delayed evaporation through the public \verb|BlackHawk| code, we evaluated the present-day Stochastic Gravitational Wave Background. We found that the delayed evaporation fundamentally reshapes the SGWB spectrum. PBHs that accrete heavily survive deeper into the universe's history; their relative energy density grows larger, and their emitted gravitons suffer less volumetric dilution. Consequently, models with initial parameters near the critical runaway limit produce explosive, multi-order-of-magnitude boosts in the GW amplitude. We showed that these signals push previously unobservable PBH parameter spaces right to the brink of the $\Delta N_{\rm eff} \le 0.17$ bounds from Planck, making them prime targets for the projected sensitivities of next-generation observatories like CMB-S4.

In summary, the assumption of a static or purely radiation-dominated early Universe vastly underestimates the complexity of PBH evolution. The intricate dance between scalar field accretion, radiation production, and cosmological expansion during the reheating epoch dictates the physical characteristics of these BHs. Accurately modeling this full thermal history is not merely a theoretical exercise, but an absolute necessity for predicting their observable signatures and utilizing them as profound probes of the early Universe.


\section*{acknowledgments}

We are grateful to Kaustubh Mukund Vispute for valuable discussions during the early stages of this work. JK would also like to thank Ayan Chakraborty for helpful discussions. The work of JK is supported by the Ministry of Human Resource Development, Government of India.

\appendix

\section{Properties and Asymptotic Expansion of the Confluent Heun Function}
\label{app:heun_function}


\paragraph{The Confluent Heun Equation:} The Confluent Heun equation is a standard differential equation with two regular singular points and one irregular singular point, defined as
\begin{equation}
    \frac{d^{2}U}{dz^{2}} + \left( \alpha + \frac{\beta+1}{z} + \frac{\gamma+1}{z-1} \right)\frac{dU}{dz} + \left( \frac{\mu}{z} + \frac{\nu}{z-1} \right) U = 0,
    \label{eq:heun_general}
\end{equation}
where the auxiliary parameters $\mu$ and $\nu$ are defined in terms of the fundamental Heun parameters $(\alpha, \beta, \gamma, \delta, \eta)$ as
\begin{align}
    \mu &= \frac{1}{2}\left(\alpha - \beta - \gamma + \alpha\beta - \beta\gamma\right) - \eta, \\
    \nu &= \frac{1}{2}\left(\alpha + \beta + \gamma + \alpha\gamma + \beta\gamma\right) + \delta + \eta.
\end{align}
The canonical solution to this equation is the Confluent Heun function, denoted as $U(z) = \mathrm{HeunC}(\alpha,\beta, \\ \gamma,\delta,\eta,z)$. To map our specific physical system to these parameters, we transform the Confluent Heun equation into its normal form using the transformation
\begin{equation}
    Z(z) = U(z) \exp\left[ \frac{1}{2} \int \left( \alpha + \frac{\beta+1}{z} + \frac{\gamma+1}{z-1} \right) dz \right].
\end{equation}
This yields the normalized equation 
\begin{equation}
    \frac{d^{2}Z}{dz^{2}} + \left[ B_{1} + \frac{B_{2}}{z^{2}} + \frac{B_{3}}{z} + \frac{B_{4}}{(z-1)^{2}} + \frac{B_{5}}{z-1} \right] Z = 0,
\end{equation}
where the coefficients $B_i$ correspond directly to the physical effective potential parameters and are given by
\begin{align}
    B_{1} &= -\frac{1}{4}\alpha^{2}, \quad
    B_{2} = \frac{1}{4}\left(1 - \beta^{2}\right), \quad
    B_{3} = \frac{1}{2}\left(1 - 2\eta\right), \\
    B_{4} &= \frac{1}{4}\left(1 - \gamma^{2}\right), \quad
    B_{5} = \frac{1}{2}\left(-1 + 2\delta + 2\eta\right). \notag
\end{align}


\paragraph{Series Solution and Asymptotic Expansion:} To extract the physical behavior in the matching region, we seek a power series solution around the regular singular point $z=0$
\begin{equation}
    U(z) = \sum_{n=0}^{\infty} a_n z^n, \qquad |z| < 1.
\end{equation}
Applying the boundary conditions $U(0) = a_0 = 1$ and evaluating the first derivative at the origin gives
\begin{equation}
    U'(0) = a_1 = \frac{\beta(1+\gamma) - \alpha(1+\beta) + 2\eta + \gamma}{2(1+\beta)}.
\end{equation}
Substituting the power series into the differential equation yields the general three-term recurrence relation
\begin{equation}
    P_n a_n = Q_n a_{n-1} + R_n a_{n-2},
\end{equation}
where the recurrence coefficients are
\begin{align}
    P_n &= n(n+\beta), \notag \\
    Q_n &= (n-1)(n+\beta+\gamma-\alpha) + \frac{\beta(1+\gamma) - \alpha(1+\beta) + 2\eta + \gamma}{2}, \\
    R_n &= (n-2)\alpha + \frac{1}{2}\alpha(\beta+\gamma+2) + \delta. \notag
\end{align}
For the $s$-wave zero-momentum mode considered in the main text, the required function is $\mathrm{HeunC} (0, \\ -2i\mu r_s, 0, -\mu^2 r_s^2, \mu^2 r_s^2, 1-r/r_s)$. This corresponds to the parameter set
\begin{equation}
    \alpha = 0, \qquad \gamma = 0, \qquad \eta = -\delta.
\end{equation}
Under these conditions, the recurrence coefficients drastically simplify to
\begin{align}
    P_n = n(n+\beta), \quad \quad
    Q_n = (n-1)(n+\beta) + \frac{\beta - 2\delta}{2}, \quad \quad
    R_n = \delta. 
\end{align}
Using the boundary condition $a_1 = (\beta - 2\delta) / [2(1+\beta)]$, we iteratively generate the higher-order terms
\begin{align}
    a_2 &= \frac{(\beta-2\delta)(3\beta-2\delta+4)}{8(\beta+1)(\beta+2)}, \notag \\
    a_3 &= \frac{(\beta-2\delta)(3\beta-2\delta+4)(5\beta-2\delta+12)}{48(\beta+1)(\beta+2)(\beta+3)}, \\
    a_4 &= \frac{(\beta-2\delta)(3\beta-2\delta+4)(5\beta-2\delta+12)(7\beta-2\delta+24)}{384(\beta+1)(\beta+2)(\beta+3)(\beta+4)}. \notag
\end{align}
Summing these terms and organizing the expansion by orders of $\beta$, $\beta^2$, and $\delta$, the function takes the analytic form
\begin{equation}
\begin{split}
    \mathrm{HeunC}(0,\beta,0,\delta,-\delta,z) &= 1 - \frac{\beta}{2} \log(1-z) + \frac{\beta^{2}}{4}\left[\log(1-z) - \operatorname{Li}_2(1-z)\right] \\
    &\quad - \frac{\delta}{2}\left(z - \log(1-z)\right) + \mathcal{O}(z^3),
\end{split}
\end{equation}
where $\operatorname{Li}_2(x)$ is the Polylog function of order 2. Finally, substituting our physical parameters $\beta = -2 i \mu r_s$, $\delta = -\mu^{2} r_s^2$, and the coordinate mapping $z = 1 - r/r_s$, the expansion becomes
\begin{equation}
\begin{split}
    & \mathrm{HeunC}\left(0, -2i\mu r_s, 0, -\mu^2 r_s^2, \mu^2 r_s^2, 1-\frac{r}{r_s}\right) \\
    & \;\;\; \simeq 1 + i \mu r_s \log \left( \frac{r}{r_s} \right) - \frac{1}{2} \mu^{2} r_s^2 \left[ \frac{r}{r_s} + 3 \log \left(\frac{r}{r_s}\right) - 2\operatorname{Li}_2 \left(1-\frac{r}{r_s}\right) - 1 \right] + \mathcal{O}(\mu^{3}r_s^3).
\end{split}
\label{eq:Heunc_approx}
\end{equation}
This expression dictates the asymptotic behavior of the scalar field necessary for matching to the background cosmological solution.

\section{Derivation of the Corrected Hubble Parameter}
\label{app:hubble_derivation}

In Section~\ref{subsec:amplitude_damping}, we presented the evolution of the Hubble parameter including small oscillations on top of the averaged background evolution. Here, we provide the detailed derivation for this result, Eq.~\eqref{eq:H_corrected}. We begin with the Friedmann equation, expressing the Hubble parameter in terms of the instantaneous energy density of the scalar field
\begin{equation}
    H^2 = \frac{1}{3\mpl^2}\rho_\Phi = \frac{1}{3\mpl^2}\left(\frac{1}{2}\dot{\Phi}^2 + V(\Phi)\right).
\end{equation}
We substitute the solution $\Phi(t) = \Phi_0(t)\mathcal{P}(t)$, where the time evolution of the amplitude is given by $\dot{\Phi}_0 = - \frac{3H}{n+1}\Phi_0$. The time derivative of the field is $\dot{\Phi} = \dot{\Phi}_0\mathcal{P} + \Phi_0 \dot{\mathcal{P}}$. Substituting these into the Friedmann equation gives~\cite{Chakraborty:2023lpr}
\begin{align}
3\mpl^2 H^2 = \frac{1}{2}\Phi_0^2\left(\dot{\mathcal{P}}^2 - \frac{6H}{n+1}\mathcal{P}\dot{\mathcal{P}} + \frac{9H^2}{(n+1)^2}\mathcal{P}^2\right) + V(\Phi_0)\mathcal{P}^{2n}.
\end{align}
This can be rearranged into a quadratic equation for the Hubble parameter, $A H^2 + B H + C = 0$, with the coefficients
\begin{align}
    A = 3\mpl^2 - \frac{9\Phi_0^2\mathcal{P}^2}{2(n+1)^2}, \quad
    B = \frac{3\Phi_0^2\mathcal{P}\dot{\mathcal{P}}}{n+1}, \quad
    C = -\frac{1}{2}\Phi_0^2\dot{\mathcal{P}}^2-V(\Phi_0)\mathcal{P}^{2n}.
\end{align}
The physically relevant (positive) solution is $H = \frac{-B + \sqrt{B^2 - 4AC}}{2A}$. The term under the square root is the most complex part. Let's analyze it, assuming $\Phi_0 \ll \mpl$, which is valid after inflation
\begin{equation}
B^2 - 4AC \approx 12\mpl^2 \left(\frac{1}{2}\Phi_0^2\dot{\mathcal{P}}^2+V(\Phi_0)\mathcal{P}^{2n}\right).
\end{equation}
Using the relations $V(\Phi_0) = \langle \rho_\Phi \rangle = 3\mpl^2 H_{\text{avg}}^2$ and the Virial theorem result $\dot{\mathcal{P}}^2 = \frac{m_\Phi^2}{n(2n-1)}(1-\mathcal{P}^{2n})$, where $m_\Phi^2 = V''(\Phi_0) =2n(2n-1)V(\Phi_0)/\Phi_0^2 $, we can simplify the term $\frac{1}{2}\Phi_0^2\dot{\mathcal{P}}^2+V(\Phi_0)\mathcal{P}^{2n} = V(\Phi_0)$. So, the term under the square root becomes $\sqrt{B^2-4AC} \approx 2\sqrt{3} \mpl \sqrt{V(\Phi_0)} $. Now, solving for $H$ gives
\begin{align}
    H \approx \sqrt{\frac{V(\Phi_0)}{3\mpl^2}} - \frac{\Phi_0^2\mathcal{P}\dot{\mathcal{P}}}{2\mpl^2(n+1)}.
\end{align}
The first term is the averaged Hubble parameter, $H_{\text{avg}} = \sqrt{V(\Phi_0)/(3\mpl^2)}$. For the second term, we use $\dot{\mathcal{P}} = \pm \frac{m_\Phi}{\sqrt{n(2n-1)}}\sqrt{1-\mathcal{P}^{2n}}$. This gives
\begin{align}
    H \approx  H_{\text{avg}} \left( 1 \mp \frac{\mathcal{P}\sqrt{6(1-\mathcal{P}^{2n})}}{2(n+1)} \frac{\Phi_0}{\mpl} \right).
\end{align}
The sign depends on the direction of oscillation. Re-writing with a conventional sign, we arrive at the final expression used in the main text
\begin{equation}
H(t) \approx H_{\text{avg}}(t) \left( 1 + \frac{\mathcal{P}\sqrt{6(1-\mathcal{P}^{2n})}}{2(n+1)} \frac{\Phi_0(t)}{\mpl} \right).
\end{equation}


\end{document}